\documentclass[acmsmall,nonacm]{acmart}

\newtoggle{extended}
\toggletrue{extended}
\iftoggle{extended}{
  \setcopyright{cc}
}{
  \setcopyright{rightsretained}
  \acmDOI{10.1145/3649835}
  \acmYear{2024}
  \copyrightyear{2024}
  \acmSubmissionID{oopslaa24main-p79-p}
  \acmJournal{PACMPL}
  \acmVolume{8}
  \acmNumber{OOPSLA1}
  \acmArticle{118}
  \acmMonth{4}
  \received{20-OCT-2023}
  \received[accepted]{2024-02-24}
}

\usepackage{booktabs}   %% For formal tables:
\usepackage{subcaption} %% For complex figures with subfigures/subcaptions
\usepackage{color}
\usepackage[textwidth=3.1cm]{todonotes}
\usepackage{xspace}
\usepackage[frozencache=true,cachedir=minted-cache]{minted}
\usepackage{cleveref}
\usepackage{mathpartir}
\usepackage{etoolbox}
\usepackage{tikz}
\usepackage{amsmath}
\usepackage{wrapfig}

\definecolor{mike}{rgb}{0.91, 0.84, 0.42}
\definecolor{cdiss}{HTML}{6ED7F7}
\definecolor{awells}{HTML}{ff3377}
\definecolor{shaobo}{HTML}{33FFBD}
\definecolor{aeline}{HTML}{22c442}

\paperwidth=\dimexpr \paperwidth + 5cm\relax
\oddsidemargin=\dimexpr\oddsidemargin + 2cm\relax
\evensidemargin=\dimexpr\evensidemargin + 2cm\relax
\marginparwidth=\dimexpr \marginparwidth + 2cm\relax

\newtoggle{todos}
\togglefalse{todos}

\iftoggle{todos}{
  \newcommand{\Todo}[3]{\todo[color=#1,size=\tiny]{#2: #3}}
}{
  \newcommand{\Todo}[3]{}
}

\iftoggle{todos}{
  \newcommand{\TODO}[1]{{\color{red}TODO: #1}}
}{
  \newcommand{\TODO}[1]{}
}
\iftoggle{todos}{
  
}{
  
}
\iftoggle{todos}{
  \newcommand{\tocite}[1]{{\color{red}[tocite: #1]}}
}{
  \newcommand{\tocite}[1]{}
}
\newcommand{\code}[1]{%
  \mintinline[fontsize=\small{},mathescape,escapeinside=@@]{go}{#1}%
}

\newminted[cedarblock]{go}{fontsize=\footnotesize{},mathescape,breaklines,escapeinside=@@}
\newminted[schemablock]{lisp}{fontsize=\footnotesize{},mathescape,breaklines,escapeinside=@@}
\newminted[fgablock]{text}{fontsize=\footnotesize{},mathescape,breaklines,escapeinside=@@}
\newminted[regoblock]{python}{fontsize=\footnotesize{},mathescape,breaklines,escapeinside=@@}

\newtheorem{example}{Example}[section]

\newcommand{\cedar}{Cedar\xspace}
\newcommand{\fga}{OpenFGA\xspace}
\newcommand{\rego}{Rego\xspace}

\newcommand{\gdrive}{\code{gdrive}\xspace}
\newcommand{\github}{\code{github}\xspace}
\newcommand{\gdrivet}{\code{gdrive-templates}\xspace}
\newcommand{\githubt}{\code{github-templates}\xspace}
\newcommand{\micros}{{$\mu$}s\xspace}

\newcommand{\tighten}{\looseness=-1}

\usepackage[english]{babel}
\addto\extrasenglish{
  
}

\newcommand{\cedarfasterthanfgagdrive}{28.7$\times$\xspace} % 134.7 µs / 4.7 µs
\newcommand{\cedarfasterthanfgagithub}{34.4$\times$\xspace} % 378.7 µs / 11.0 µs
\newcommand{\cedarfasterthanfgatinytodo}{35.2$\times$\xspace} % 175.8 µs / 5.0 µs
\newcommand{\cedarfasterthanfga}{28.7$\times$-35.2$\times$\xspace}
\newcommand{\cedarfasterthanregogdrive}{60.4$\times$\xspace} % 284.0 µs / 4.7 µs
\newcommand{\cedarfasterthanregogithub}{80.8$\times$\xspace} % 888.9 µs / 11.0 µs
\newcommand{\cedarfasterthanregotinytodo}{42.8$\times$\xspace} % 213.8 µs / 5.0 µs
\newcommand{\cedarfasterthanrego}{42.8$\times$-80.8$\times$\xspace}
\newcommand{\cedarpolicyslicingfaster}{10.0$\times$-18.0$\times$\xspace} % 93.0 µs / 9.3 µs (github templates w/ and w/o slicing) ; 163.8 µs / 9.1 µs (gdrive templates w/ and w/o slicing)

\begin{document}

%% Title information
\iftoggle{extended}{
  \title{\cedar: A New Language for Expressive, Fast, Safe, and Analyzable Authorization (Extended Version)}
}{
  \title{\cedar: A New Language for Expressive, Fast, Safe, and Analyzable Authorization}
}

%% Author information
%% Contents and number of authors suppressed with 'anonymous'.
%% Each author should be introduced by \author, followed by
%% \authornote (optional), \orcid (optional), \affiliation, and
%% \email.
%% An author may have multiple affiliations and/or emails; repeat the
%% appropriate command.
%% Many elements are not rendered, but should be provided for metadata
%% extraction tools.

\author{Joseph W. Cutler}
\authornote{Work carried out while at Amazon Web Services}          %% \authornote is optional;
%\orcid{nnnn-nnnn-nnnn-nnnn}             %% \orcid is optional
\affiliation{
  \institution{University of Pennsylvania}
  \country{USA}
}
\email{jwc@seas.upenn.edu}
\orcid{0000-0001-9399-9308}
%%%%
\author{Craig Disselkoen}
\affiliation{
  \institution{Amazon Web Services}
  \country{USA}
}
\email{cdiss@amazon.com}
\orcid{0000-0003-4358-2963}
%%%%
\author{Aaron Eline}
\affiliation{
  \institution{Amazon Web Services}
  \country{USA}
}
\email{aeline@amazon.com}
\orcid{0000-0002-9105-4922}
%%%%
\author{Shaobo He}
\affiliation{
  \institution{Amazon Web Services}
  \country{USA}
}
\email{shaobohe@amazon.com}
\orcid{0000-0002-9899-6226}
%%%%
\author{Kyle Headley}
\authornotemark[1]
\affiliation{
  \institution{Unaffiliated}
  \country{USA}
}
\email{kylenheadley@gmail.com}
\orcid{0000-0002-4880-4150}
%%%%
\author{Michael Hicks}
\affiliation{
  \institution{Amazon Web Services}
  \country{USA}
}
\email{mwhicks@amazon.com}
\orcid{0000-0002-2759-9223}
%%%%
\author{Kesha Hietala}
\affiliation{
  \institution{Amazon Web Services}
  \country{USA}
}
\email{khieta@amazon.com}
\orcid{0000-0002-2724-0974}
%%%%
\author{Eleftherios Ioannidis}
\authornotemark[1]
\affiliation{
  \institution{University of Pennsylvania}
  \country{USA}
}
\email{elefthei@seas.upenn.edu}
\orcid{0000-0003-2749-797X}
%%%%
\author{John Kastner}
\affiliation{
  \institution{Amazon Web Services}
  \country{USA}
}
\email{jkastner@amazon.com}
\orcid{0000-0002-1273-5990}
%%%%
\author{Anwar Mamat}
\authornotemark[1]
\affiliation{
  \institution{University of Maryland}
  \country{USA}
}
\email{anwar@umd.edu}
\orcid{0009-0007-1184-7206}
%%%%
\author{Darin McAdams}
\affiliation{
  \institution{Amazon Web Services}
  \country{USA}
}
\email{darinm@amazon.com}
\orcid{0009-0002-4005-1817}
%%%%
\author{Matt McCutchen}
\authornotemark[1]
\affiliation{
  \institution{Unaffiliated}
  \country{USA}
}
\email{matt@mattmccutchen.net}
\orcid{0000-0003-4814-5148}
%%%%
\author{Neha Rungta}
\affiliation{
  \institution{Amazon Web Services}
  \country{USA}
}
\email{rungta@amazon.com}
\orcid{0000-0001-5143-8940}
%%%%
\author{Emina Torlak}
\affiliation{
  \institution{Amazon Web Services}
  \country{USA}
}
\email{torlaket@amazon.com}
\orcid{0000-0002-1155-2711}
%%%%
\author{Andrew M. Wells}
\affiliation{
  \institution{Amazon Web Services}
  \country{USA}
}
\email{anmwells@amazon.com}
\orcid{0000-0001-7780-2122}

% Joseph Cutler (University of Pennsylvania) <jwc@seas.upenn.edu>
% Craig Disselkoen (Amazon Web Services) <cdiss@amazon.com>
% Aaron Eline (Amazon Web Services) <aeline@amazon.com>
% Shaobo He (Amazon Web Services) <shaobohe@amazon.com>
% Kyle Headley (Unaffiliated) <kylenheadley@gmail.com>
% Michael Hicks (Amazon Web Services) <mwhicks@amazon.com>
% Kesha Hietala (Amazon Web Services) <khieta@amazon.com>
% Elefterios Ioannidis (University of Pennsylvania) <elefthei@seas.upenn.edu>
% John Kastner (Amazon Web Services) <jkastner@amazon.com>
% Anwar Mamat (University of Maryland) <anwar@umd.edu>
% Darin McAdams (Amazon Web Services) <darinm@amazon.com>
% Matt McCutchen (Unaffiliated) <matt@mattmccutchen.net>
% Neha Rungta (Amazon Web Services) <rungta@amazon.com>
% Emina Torlak (Amazon Web Services) <torlaket@amazon.com>
% Andrew Wells (Amazon Web Services) <anmwells@amazon.com>

%% Abstract
%% Note: \begin{abstract}...\end{abstract} environment must come
%% before \maketitle command

\begin{abstract}
\cedar is a new authorization policy language designed to be ergonomic, fast, safe, and analyzable.
Rather than embed authorization logic in an application's code, developers can write that logic as \cedar policies and delegate access decisions to \cedar's evaluation engine.
\cedar's simple and intuitive syntax supports common authorization use-cases with readable policies, naturally leveraging concepts from role-based, attribute-based, and relation-based access control models.
\cedar's policy structure enables access requests to be decided quickly.
\cedar's policy validator leverages optional typing to help policy writers avoid mistakes, but not get in their way.
\cedar's design has been finely balanced to allow for a sound and complete logical encoding, which enables precise policy analysis, e.g., to ensure that when refactoring a set of policies, the authorized permissions do not change.
We have modeled \cedar in the Lean programming language, and used Lean's proof assistant to prove important properties of \cedar's design.
We have implemented \cedar in Rust, and released it open-source.
Comparing \cedar to two open-source languages, \fga and \rego, we find (subjectively) that \cedar has equally or more readable policies, but (objectively) performs far better.
\end{abstract}

\iftoggle{extended}{

}
{
  %   <ccs2012>
  %      <concept>
  %          <concept_id>10002978.10002986</concept_id>
  %          <concept_desc>Security and privacy~Formal methods and theory of security</concept_desc>
  %          <concept_significance>300</concept_significance>
  %          </concept>
  %      <concept>
  %          <concept_id>10003752.10010124</concept_id>
  %          <concept_desc>Theory of computation~Semantics and reasoning</concept_desc>
  %          <concept_significance>300</concept_significance>
  %          </concept>
  %      <concept>
  %          <concept_id>10002978.10002991.10010839</concept_id>
  %          <concept_desc>Security and privacy~Authorization</concept_desc>
  %          <concept_significance>500</concept_significance>
  %          </concept>
  %    </ccs2012>
  %   \end{CCSXML}


}

\maketitle
\renewcommand{\shortauthors}{Cutler et al.}

\section{Introduction}
\label{sec:intro}

\newcommand{\challenge}[1]{\textbf{#1.}}

Authorization is the problem of deciding who has access to what in a multi-user system.
Every cloud-based application has to solve this problem, from photo sharing to online banking to health care.
A common solution is to embed access control logic in the application code, as shown in the \code{get_list} method for TinyTodo (\autoref{fig:embedded-access-control}), a hypothetical application for managing todo lists.

\begin{wrapfigure}{r}{.77\textwidth}
\begin{minted}[fontsize=\footnotesize{},linenos,xleftmargin=2em]{python}
def get_list(request):
    if not db.query(request.user).admin:
        if db.query(request.listId).owner != request.user:
            if not request.user in db.query(request.listId).readers:
                if not request.user in db.query(request.listId).editors:
                    return 'AccessDenied'
    list = db.query(request.listId)
    return { 'id': list.id, 'owner': list.owner, ... }
\end{minted}
\caption{Embedded access control logic in the \code{get_list} method for TinyTodo\tighten}\label{fig:embedded-access-control}
\end{wrapfigure}

This approach has three major drawbacks.
First, it is \emph{hard to write} correct permissions.
The logic in \code{get_list} denies access if the requester is neither an admin, nor the list's owner, nor a member of its readers or writers groups.
It would be easy to mistakenly drop a \code{not} or improperly nest the conditions; such mistakes account for four of the top 25 security weaknesses in 2023~\cite{cwe2023top25}.
Second, embedded permissions are \emph{hard to understand and audit}.
An auditor needs to read the application's code to check access invariants such as \emph{a list's editors can perform all the same actions as the list's readers}.
Finally, embedded permissions are \emph{hard to maintain}.
Changing authorization logic means changing the application code, so policy versioning is tantamount to code versioning.
Adding crosscutting permissions (e.g., supporting a new user role) requires updates to code in multiple places, which offers more chances for mistakes.\tighten

\challenge{Policies as code}
A better alternative is to externalize access control rules into policies written in a domain-specific \emph{authorization language}, and delegate decisions to the language's \emph{authorization engine}.
This approach is called \emph{policies as code}~\cite{policiescode}.
With policies as code, lines 2--6 of \code{get_list} become a single, unchanging call to the authorization engine, e.g.,
\mintinline[fontsize=\footnotesize{},mathescape,escapeinside=@@]{python}{if not client.is_authorized(...): }
\mintinline[fontsize=\footnotesize{},mathescape,escapeinside=@@]{python}{return 'AccessDenied'}.
The authorization engine makes a decision by consulting the application's policies, which are expressed separately from the application code, in a dedicated DSL.
This makes access control rules easier to understand, audit, change, version, and share between applications.\tighten

But building a high-quality authorization language is a significant challenge that requires balancing four competing goals.
In particular, the language should be
\textbf{expressive}, so that application developers can grant access based on user and resource attributes, group membership, and session context~\cite{rbac,abac,zanzibar2019};
\textbf{performant}, so that requests are decided quickly;
\textbf{safe}, so that policy authoring mistakes are caught or avoided; and
\textbf{analyzable}, so that policy analysis tools can reason precisely about important access invariants.\tighten

The first goal is in tension with the other three.
A highly expressive language may have constructs that are slow to execute, hard to prove safe, and impossible to analyze precisely.
A fast, safe, and analyzable language, in contrast, may be too weak to express common authorization use-cases.
To our knowledge (\Cref{sec:related}), no existing language finds the ideal balance of these four goals.\tighten

\challenge{\cedar: a novel authorization language}
This paper presents \cedar, a new authorization language that is simultaneously expressive, performant, safe, and analyzable.
\cedar is used at scale in products and services from Amazon Web Services, and is open source.

\cedar's vocabulary can naturally \textbf{express} who has access to what based on users' roles and attributes of their environment.
Most \cedar operators take constant time, and looping constructs are linear, ensuring \textbf{fast} policy evaluation times.
While deciding a request considers all active policies, \cedar is able to quickly \emph{slice} the full policy set to evaluate only those relevant to the request.\tighten

\cedar's design creates a \textbf{safe} foundation for authoring policies.
Policies have no side effects, and decisions are indifferent to the order that policies are considered.
\cedar policies never authorize an action by default---it must be explicitly permitted.
\cedar provides a \emph{policy validator}, based on a novel type system leveraging \emph{singleton types}~\cite{aspinall94singleton} and \emph{static capabilities}~\cite{capabilities1999}, that ensures policy evaluation will never error on a type mismatch, or a bogus attribute or role name.\tighten

\cedar's design ensures that its policies are efficiently \textbf{analyzable} by reduction to SMT, so that access invariants can be proved automatically.
% For example, by defining the semantics of \texttt{in} on the transitive closure rather than on the entity DAG, \cedar's \emph{symbolic compiler} can avoid the use of an undecidable transitive-closure logic~\cite{tclogics2004}.
We define a novel \emph{symbolic compiler} that translates \cedar policies to SMT, producing a decidable, sound, and complete encoding of their semantics.\tighten

\challenge{Implementation and Evaluation}
We have formalized \cedar in Lean~\cite{lean4} and used Lean's proof assistant to prove the properties mentioned above.
We have implemented \cedar in Rust, and used extensive differential random testing to ensure that the implementation matches the model.

We conduct three sets of experiments to evaluate \cedar's expressiveness, performance, and analyzability.
First, we use \cedar to encode the authorization models for two example applications previously modeled in \fga~\cite{openFGA},
%\rego is a popular authorization language based on Datalog, and \fga is
a recently developed language for specifying relationship-based access control (ReBAC)~\cite{zanzibar2019} policies.
We can express both of these \fga models with validated \cedar policies.
Next, we compare the performance of \cedar, \fga, and \rego~\cite{opa}---a popular authorization language based on Datalog---on three example applications: the two \fga examples and an additional example described in \Cref{sec:overview}.
Comparing the performance of these three models on randomly generated inputs, we find that the \cedar authorizer is \cedarfasterthanfga faster than \fga and \cedarfasterthanrego faster than \rego.
We also evaluate the effectiveness of \cedar's policy slicing, and find that authorization is \cedarpolicyslicingfaster faster on average, when using a \cedar templates-based encoding.
Finally, we find that \cedar's type system is powerful enough to validate all policies in our example models, and \cedar's symbolic compiler reduces analysis questions about these policies into SMT-LIB formulas that take, on average, 75.1 ms to encode and solve.\tighten

\smallskip

In summary, this paper makes the following contributions:%
\iftoggle{extended}{
\footnote{This paper extends a peer-reviewed publication of the same name~\cite{cedar-oopsla} with an appendix of policy examples.}
}{}
\begin{itemize}
  \item Design, implementation, and evaluation of \cedar, a new authorization language that balances expressiveness, safety, performance, and analyzability.\tighten
  \item A verified validator for \cedar, designed to accept safe policies that are translatable to SMT.\tighten
  \item A verified symbolic compiler for reducing \cedar to a decidable fragment of SMT.\tighten
\end{itemize}
\cedar's code, documentation, and example apps can be found at \url{https://github.com/cedar-policy/}.

% Use the TinyTodo blog post (\url{https://aws.amazon.com/blogs/opensource/using-open-source-cedar-to-write-and-enforce-custom-authorization-policies/}) and tutorial (\url{https://github.com/cedar-policy/cedar-examples/blob/main/tinytodo/TUTORIAL.md}) as the basis for this, demo'ing \cedar syntax, data model, validation.

% local defines for this section, just to save some typing and ensure consistency
\newcommand{\kw}[1]{\ensuremath{\mathtt{#1}}}
\newcommand{\principal}{\kw{principal}}
\newcommand{\action}{\kw{action}}
\newcommand{\resource}{\kw{resource}}
\newcommand{\context}{\kw{context}}
\newcommand{\ty}[1]{\ensuremath{\mathit{#1}}}
\newcommand{\tybool}{\ty{Bool}}
\newcommand{\tystring}{\ty{String}}
\newcommand{\tylong}{\ty{Long}}
\newcommand{\tytrue}{\ty{True}}
\newcommand{\tyfalse}{\ty{False}}
\newcommand{\tyset}[1]{\ty{Set}~{#1}}
\newcommand{\etrue}{\kw{true}}
\newcommand{\efalse}{\kw{false}}

\newcommand{\user}{\code{User}\xspace}
\newcommand{\users}{\code{User}s\xspace}
\newcommand{\team}{\code{Team}\xspace}
\newcommand{\teams}{\code{Team}s\xspace}
\newcommand{\tasklist}{\code{List}\xspace}
\newcommand{\tasklists}{\code{List}s\xspace}

\section{Overview}
\label{sec:overview}

We overview \cedar using TinyTodo, the application introduced in the prior section.
We illustrate the structure and semantics of \cedar policies and its data model, how the \cedar validator typechecks \cedar policies to prevent run-time errors, and how the \cedar symbolic compiler allows us to reason about the meaning of \cedar policies.\tighten

\subsection{Basic TinyTodo policies: Creating and managing lists}

TinyTodo allows individuals, called \users, and groups, called \teams, to organize, track, and share their todo \tasklists.
% \users create \tasklists which they can populate with tasks.
% As tasks are completed, they can be checked off the list.
% TinyTodo users should not be able to see or make changes to just any task list.
We specify and enforce TinyTodo's access rules using \cedar policies.

Consider the first two of TinyTodo's policies shown in \Cref{fig:tinytodo-policies}.
Policy 0 states that any principal (a TinyTodo \user) can create a list or get a listing of their previously created lists.
The resource \code{Application::"TinyTodo"} represents the TinyTodo application itself, which is a container for any created list.
Policy 1 states that any principal can perform any action on a resource that is a \tasklist and whose \code{owner} attribute matches the requesting principal.
TinyTodo always sets a \tasklist's \code{owner} at the time it is created to the \user that created it.\tighten

Suppose there is a TinyTodo user \code{andrew} who attempts to create a new todo list called \texttt{"Create demo"}, add two tasks to that list, and then complete one of the tasks.
Creating the list is authorized by Policy 0: Any user is allowed to create a \tasklist.
The other three actions are authorized by Policy 1:
since \code{andrew} is the owner of the \tasklist, he may carry out any action on it.

\begin{figure}
\centering
\begin{tabular}{ll}
\begin{minipage}{.5\textwidth}
\begin{cedarblock}
// Policy 0: Any User can create a list
// and see what lists they own.
permit(
    principal,
    action in [Action::"CreateList",
               Action::"GetOwnedLists"],
    resource == Application::"TinyTodo");

// Policy 2: Admins can perform any action.
permit(
    principal in Team::"admin",
    action,
    resource in Application::"TinyTodo");

// Policy 4: Interns can't create task lists.
forbid(
    principal in Team::"interns",
    action == Action::"CreateList",
    resource == Application::"TinyTodo");
\end{cedarblock}
\end{minipage} &
\begin{minipage}{.47\textwidth}
\begin{cedarblock}
// Policy 1: Any User can perform any action
// on a List they own.
permit(principal, action, resource)
when {
    resource is List &&
    resource.owner == principal
};

// Policy 3: A User can see a List
// if they are either a reader or editor.
permit(
    principal,
    action == Action::"GetList",
    resource)
when {
    principal in resource.readers ||
    principal in resource.editors
};
\end{cedarblock}
\end{minipage} \\
\end{tabular}
\caption{\cedar policies for TinyTodo}
\label{fig:tinytodo-policies}
\end{figure}

\subsection{\cedar data model: Hierarchical entities}
\label{sec:data-model}

Principals, resources, and actions are represented by \cedar \emph{entities}.
Entities are collected in an \emph{entity store} and referenced by a unique identifier consisting of two parts: the \emph{entity type} and \emph{entity ID}.
Each entity is associated with zero or more attributes mapped to values, and zero or more parent entities.
The parent relation on entities forms a directed acyclic graph (DAG), called the \emph{entity hierarchy}.\tighten

\begin{figure}
    \includegraphics[width=.8\textwidth]{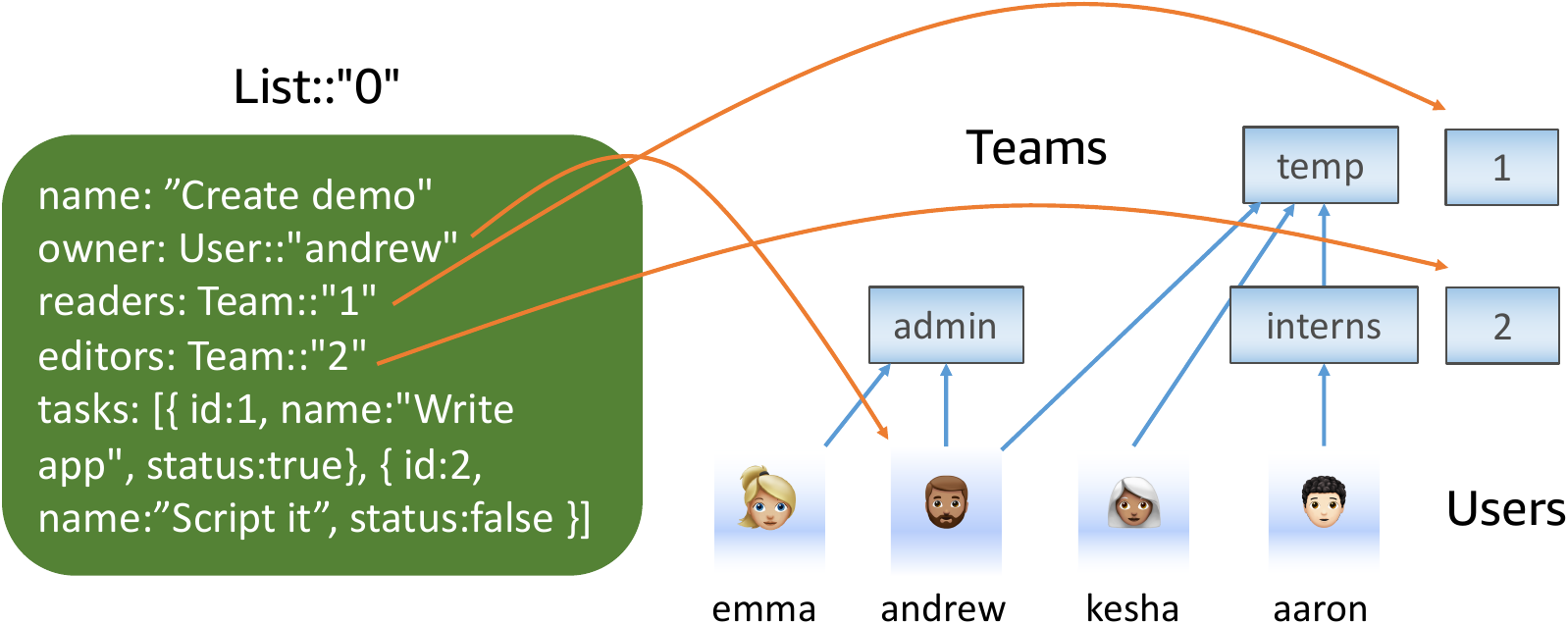}
    \caption{Example \cedar entity hierarchy for TinyTodo}
    \label{fig:hierarchy-example}
\end{figure}

\Cref{fig:hierarchy-example} illustrates the TinyTodo entities after carrying out the actions described above.
On the left is the entity for the list \code{andrew} created, which has entity reference \code{List::"0"}.
% Here, \code{List} is the entity's type, and \code{"0"} is its ID\@.
We can see that the entity has five attributes, \code{name}, \code{owner}, \code{readers}, \code{editors}, and \code{tasks}.
All entities of type \code{List} have these attributes.
On the right of the figure we see four \code{User}-typed entities (along the bottom) and five \code{Team}-typed entities (along the top), which are arranged in a hierarchy; each arrow points to an entity's immediate parent.
% For example, the entity with reference \code{User::"andrew"} has parent \code{Team::"admin"}, and entity \code{User::"aaron"} has parent \code{Team::"interns"} which in turn has parent \code{Team::"temp"}.
% The values of \code{List::"0"}'s attributes include references to other entities.
TinyTodo set these attributes, and created \code{Team::"1"} and \code{Team::"2"}, when it created the list, and updated \code{tasks} as tasks were added and completed.
% Notably, \code{owner} is \code{User::"andrew"}, and \code{readers} and \code{editors} are \code{Team::"1"} and \code{Team::"2"}, respectively.
% The \code{tasks} part of the list was updated when \code{andrew} added tasks to the list, and updated their status.

% The application maps its data to corresponding \cedar entities when invoking the \cedar authorizer.
An application might store its data in a SQL database or a key-value store in a format that best suits its purposes.
When it goes to authorize a user request, it provides the \cedar authorizer access to the necessary parts of that data, mapped to \cedar entities.
It could do so by mapping its native-stored data to \cedar entities at the time of the request, or during the application's operation.

\subsection{\cedar policy structure: Supporting RBAC, ABAC, ReBAC}

In general, \cedar policies have three components: the \emph{effect}, the \emph{scope}, and the \emph{conditions}.
The effect is either \code{permit} or \code{forbid}, indicating whether the policy is authorizing access or taking it away.
The scope comes after the effect, constraining the principal, action, and resource components of requests that the policy applies to.
The conditions are optional expressions that come last, adding further constraints, oftentimes based on attributes of request elements.
Policies access request elements using the variables \code{principal}, \code{action}, \code{resource}, and \code{context}.
The \code{context} record stores application-specific data about the request, such as current time or source IP address.\tighten

\cedar policies support role-based, attribute-based, and relation-based access control (RBAC, ABAC, and ReBAC, respectively).
RBAC is possible because of \cedar's support for a hierarchy of entities.
For example, Policy 2 in \Cref{fig:tinytodo-policies} is an RBAC-style policy, which states that any principal who is a member of the \code{admin} \team (such as \code{User::"emma"}) can perform any action on a TinyTodo resource.
% According to this policy \code{User::"emma"}, who is defined to be a member of \code{Team::"admin"} in \Cref{fig:hierarchy-example}, can view and edit any list.

ABAC-style policies use policy conditions to reference \emph{attributes}.
For example, Policy 1 mentions \code{resource.owner}, where \code{owner} is a attribute of \code{resource}.
As shown in \Cref{fig:hierarchy-example}, attributes can be primitive-typed values (like a number or string), collections, records, or entity references.

ReBAC-style policies use attributes that reference entity groups to express \emph{relations} between two entities.
For example, Policy 3 is a ReBAC-style policy, which states that any principal can read the contents of a task list (\code{Action::"GetList"}) if they are in either the list's \code{readers} or \code{editors} groups.
Here, \code{principal in resource.readers} and \code{principal in resource.editors} are expressions that can be viewed as querying whether \code{principal} is in the \emph{readers} and \emph{editors} relations with \code{resource}.

For our running example, suppose \code{User::"andrew"} shares \code{List::"0"} with the team \code{interns} as a reader (which he is allowed to do because of Policy 1).
As a result, TinyTodo will update the hierarchy in \Cref{fig:hierarchy-example} to add a parent edge from \code{Team::"interns"} to \code{Team::"1"}.
Now if \code{User::"aaron"} attempts to read the contents of \code{List::"0"} he will be allowed to do so according to Policy 3: its condition \code{principal in resource.readers} is true since the \code{readers} attribute of \code{List::"0"} corresponds to \code{Team::"1"}, which is an ancestor of \code{User::"aaron"} in the entity hierarchy.

\subsection{Safety}
\label{sec:deny}

\cedar has a design goal of creating a \emph{safe} foundation for writing policies.
One aspect of this goal is that \cedar's authorizer \emph{denies by default}: If no \code{permit} policy exists that authorizes a request, the request will be denied.
% Suppose that \code{User::"kesha"} attempted to access \code{List::"0"}.
% Considering all of the policies in \Cref{fig:tinytodo-policies}, this request would be denied: Her request is not explicitly forbidden by Policy 4, and not explicitly permitted by any of the other policies.\footnote{Why? Because: (1) she is not trying to create a list or list her lists, so Policy 0 does not apply; (2) she is not the \code{owner} of \code{List::"0"} so Policy 1 does not apply; (3) she is not a member of \code{Team::"admin"} so Policy 2 does not apply; (4) she is not part of \code{List::"0"}'s readers or editors groups, so Policy 3 does not apply.}

Another aspect is that \emph{\code{forbid} policies always override \code{permit} policies} when making the final decision.
For example, Policy 4 expresses that no intern is allowed to create a new task list (\code{Action::"CreateList"}).
With this policy, if \code{aaron} tries to create a list, his request will be approved by Policy 0, but denied by Policy 4, with the final decision that the request is denied.
This behavior allows \code{forbid} policies to act like guardrails that enforce universal non-access rules.

The third aspect of safety is that \cedar's authorizer is \emph{deterministic}: It is guaranteed to terminate and always produce the same authorization decision for a given request, hierarchy, and set of policies.
Because \cedar policies are free of side effects and general loops, policy evaluation order doesn't matter.
Because \cedar's algorithm for selecting the policies \emph{relevant} to a request is \emph{sound}, the authorizer will produce the same outcome as if had it considered all available policies.

\subsection{Validating policies}
\label{sec:validation-overview}

A final aspect of safety is ensuring that \cedar policy evaluation will not result in an error, e.g., if the policy attempts to access a non-existent attribute.
Policy writers can use the \cedar \emph{validator} to ensure that policies are (mostly) error-free.
To do so, they provide a \emph{schema} that lists the names and the type structure of an application's entities, as well as the application actions and the allowed shapes of requests that contain them.

\begin{figure}
\begin{schemablock}
      entity Application;
      entity Team, User in [Team, Application];
      entity List in [Application] { readers: Team, editors: Team, owner: User };
      action CreateList appliesTo { principal: [User], resource: [Application] };
      action GetList appliesTo { principal: [User], resource: [List] };
\end{schemablock}
\caption{TinyTodo schema}
\label{fig:tinytodo-schema}
\end{figure}

A schema for TinyTodo is shown in \Cref{fig:tinytodo-schema} (some actions and entity attributes are elided).
The schema declares four entity types: \code{Application}, \code{User}, \code{Team}, and \code{List}.
The \code{Application} entity type is the simplest possible: it has no associated attributes and has no parents in the entity hierarchy.
The combined \code{User} and \code{Team} entity type declaration indicates that both can have \code{Team} entities as parents (as we see in \Cref{fig:hierarchy-example}) as well as \code{Application} entities.
A \code{List} entity can have \code{Application}-entity parents and must have the attributes \code{owner}, \code{readers}, and \code{writers}.
The schema also declares two actions, \code{CreateList} and \code{GetList}, and specifies that they \emph{apply to} specific principal and resource entity types---any request involving these actions must include \code{principal} and \code{resource} entities of these types.
All of the policies in \Cref{fig:tinytodo-policies} are valid with respect to this schema.

Policy validation works by typechecking the policy specialized to each of the allowed actions in the \emph{request environment}s (maps from variables like \principal{} and \resource{} to types) induced by those actions' schemas.
% The example schema effectively defines two request type environments.
% The first, call it $\Gamma_{\mathit{CreateList}}$, is for \code{Action::"CreateList"} requests, and maps \principal{} to \code{User} and \resource{} to \code{Application}.
% The second, $\Gamma_{\mathit{GetOwnedLists}}$, is for \code{Action::"GetList"} requests, and maps \principal{} to \code{User} and \resource{} to \code{List}.
A policy is valid if it typechecks in every environment.
The type system uses a novel combination of \emph{static capabilities}~\cite{capabilities1999} and boolean \emph{singleton types}~\cite{aspinall94singleton}.
Capabilities ensure that a policy always checks the presence of an optional attribute before accessing it.
Singleton types $\tytrue$ and $\tyfalse$ are ascribed to expressions sure to evaluate to $\etrue$ and $\efalse$, respectively.
The type system uses them to type expressions like \code{false && (1 == "hello")}: since the first conjunct has type $\tyfalse$ the whole expression will---the erroring second conjunct will be short-circuited.
Singleton types are also used to warn when a policy will evaluate to \code{false} (or \code{true}) in \emph{every} environment, making it effectively useless (or over-influential).

Typechecking in each relevant environment, rather than a single generic environment, adds cost but avoids many false alarms.
There are typically few \code{principal} types and a recommended best practice is to specialize each action to a resource type (e.g., \code{Action::"GetList"}), so validation time tends to be $O(na)$ where $n$ is the size of the policies and $a$ is the number of actions.

\subsection{Analyzing policies}

The validator can flag certain mistakes, but it cannot answer deeper questions about policy behavior.
One such question is policy equivalence---do two (sets of) policies evaluate to \code{true} on the same set of requests?
For that, we turn to \cedar's \emph{symbolic compiler}, which works by reducing policies to SMT formulas, and discharging questions about their behavior using an SMT solver.

To illustrate, suppose that we drop Policy 2 and replace Policy 0 in \Cref{fig:tinytodo-policies} with Policy 0.1 shown to the right.
Is the resulting policy set (Policy 0.1 together with Policies 1 and 3) equivalent to the original (Policies 0, 1, 3 and 4)?
The answer is no: the new policy set accepts fewer requests than the original.
% Are they going to produce the same authorization decision on all possible requests and entity stores that conform to the TinyTodo schema?
\tighten

\begin{wrapfigure}{r}{.47\textwidth}
\begin{cedarblock}
    // Policy 0.1
    permit(
        principal,
        action in [Action::"CreateList",
                   Action::"GetOwnedLists"],
        resource == Application::"TinyTodo")
    unless {
        principal in Team::"interns"
    };
\end{cedarblock}
\end{wrapfigure}
Equivalence analysis uses the symbolic compiler to generate an SMT formula that is satisfiable if some request is allowed by one policy set but not the other.
If the solver finds a model for this formula, the analysis turns it into a concrete counterexample---a request and entity store---on which the two policy sets differ.
In this case, given the entity store in \Cref{fig:hierarchy-example}, the original policy set will let \code{aaron} perform the \code{GetOwnedLists} action but the new one will not.
This is because Policy 0.1 prevents interns from performing either a \code{CreateList} or \code{GetOwnedLists} action.
Adding \code{&& action == Action::"CreateList"} to the condition of Policy 0.1 fixes the problem.

We designed \cedar specifically to support an SMT encoding that is sound, decidable, and complete; no prior authorization language enjoys such an encoding.
To achieve it, we use a novel type-based translation that employs only decidable theories, and finite sets of \emph{ground} well-formedness constraints (e.g., to ensure entity graphs are acyclic) rather than \emph{quantified} constraints.
\tighten

\section{Syntax, Semantics, and Typing}
\label{sec:semantics}

This section presents a partial formalization of the syntax of \cedar policies (\Cref{sec:syntax}), the semantics of \cedar expression evaluation (\Cref{sec:expr-semantics}) and authorization (\Cref{sec:authz-semantics}), and the semantics of policy validation (\Cref{sec:types-policy}).
We have proved several properties of \cedar's design (\Cref{sec:properties})
using a formalization in Lean~\cite{lean4}.
Details about full \cedar are given in \Cref{sec:full-cedar}.

\newcommand{\eref}[2]{\ensuremath{{#1}\mathtt{::}{#2}}}
\newcommand{\ebinop}[3]{{#2}~{#1}~{#3}}
\newcommand{\ehas}[2]{\ebinop{\kw{has}}{#1}{#2}}
\newcommand{\eis}[2]{\ebinop{\kw{is}}{#1}{#2}}
\newcommand{\eand}[2]{\ebinop{\kw{\&\&}}{#1}{#2}}
\newcommand{\eor}[2]{\ebinop{\kw{||}}{#1}{#2}}
\newcommand{\econd}[3]{\kw{if}~{#1}~\kw{then}~{#2}~\kw{else}~{#3}}
\newcommand{\dom}[1]{\ensuremath{\mathit{dom}(#1)}}
\newcommand{\step}{\ensuremath{\longrightarrow}}
\newcommand{\eval}{\ensuremath{\Downarrow}}
\newcommand{\mapping}[2]{\ensuremath{{#1}\!:\!{#2}}}

\subsection{Syntax: Policies and expressions}
\label{sec:syntax}

\newcommand{\pkw}[1]{\ensuremath{\mathbf{#1}}}
\newcommand{\pcode}[1]{\ensuremath{\mathrm{#1}}}
\newcommand{\nt}[1]{\textcolor{blue}{\textit{#1}}}
\newcommand{\term}[1]{\textcolor{red}{#1}}

\begin{figure}
\begin{tabular}{l@{~~}c@{~~}ll@{~~}c@{~~}l}
\nt{Policies} ($C$) & ::= & \{~\nt{Policy}~\} &
\nt{Policy} ($c$) & ::= & \nt{Effect} \term{\kw{(}} \nt{Scope} \term{\kw{)}} \{~\nt{Cond}~\} \term{\kw{;}}\\
\nt{Effect} & ::= & \term{\kw{permit}} $\mid$ \term{\kw{forbid}} &
\nt{Scope} & ::= & \nt{Principal} \term{\kw{,}} \nt{Action} \term{\kw{,}} \nt{Resource} \\
\nt{Principal} & ::= & \term{\principal}~[(\term{\kw{in}} | \term{\kw{==}}) $\eref{E}{s}$] &
\nt{Action} & ::= & \term{\action}~[((\term{\kw{in}} $\mid$ \term{\kw{==}}) $\eref{E}{s}$) $\mid$ (\term{\kw{in}} \term{\kw{[}}$\eref{E}{s}$\term{,} ...\term{\kw{]}})] \\
\nt{Resource} & ::= & \term{\resource}~[(\term{\kw{in}} | \term{\kw{==}}) $\eref{E}{s}$] &
\nt{Cond} & ::= & (\term{\kw{when}} $\mid$ \term{\kw{unless}}) \term{\kw{\{}} $e$ \term{\kw{\}}} \\ \\
\end{tabular}

\begin{displaymath}
    \begin{array}{lccl}
        \text{Entity type} & E & \in & \mathbf{ID} \qquad \text{Attribute}~f~\in~\mathbf{ID} \qquad \text{String}~~s \qquad \text{Integer}~~i \qquad \text{Nat}~n \\[1ex]
        \text{Variable} & x & ::= & \principal \mid \action \mid \resource \mid \context \\[1ex]
        \text{Value} & v & ::= & \eref{E}{s} \mid \etrue \mid \efalse \mid s \mid i \mid [ v_1, ..., v_n ] \mid \{ f_1\!:\!v_1, ..., f_n\!:\!v_n \} \\[1ex]
        \text{Expression} & e & ::= & v \mid x \mid [ e_1, ..., e_n ] \mid \{ f_1\!:\!e_1, ..., f_n\!:\!e_n \} \mid e.f \mid \eand{e_1}{e_2} \mid \eor{e_1}{e_2} \mid \kw{!}e \mid -e \\
        && \mid & \ebinop{bop}{e_1}{e_2} \mid \ebinop{\kw{like}}{e}{s} \mid \ehas{e}{f} \mid \eis{e}{E} \mid \ebinop{\kw{*}}{i}{e} \mid \econd{e_1}{e_2}{e_3} \\[1ex]
        \text{Binop} & bop & ::= & + \mid - \mid ~<~ \mid ~\leq~ \mid \kw{==} \mid \kw{in} \mid \kw{contains} \mid \kw{containsAny} \mid \kw{containsAll} \\[1ex]
        \text{Type} & \tau & ::= & E \mid \tybool \mid \tystring \mid \tylong \mid \tytrue \mid \tyfalse \mid \tyset{\tau} \mid \{ \mapping{\omega_1 f_1}{\tau_1}, ..., \mapping{\omega_n f_n}{\tau_n} \}\\[1ex]
        \text{Optionality} & \omega & ::= & \cdot \mid ~? \qquad \qquad \qquad
        \text{Capability} ~~\alpha, \varepsilon ~~::=~~ \emptyset \mid \{ e.f \} \mid \varepsilon \cup \varepsilon \mid \varepsilon \cap \varepsilon \\
    \end{array}
    \end{displaymath}
\caption{\cedar policies, expressions, and types: Syntax}
\label{fig:syntax}
\end{figure}

The concrete syntax of \cedar policies follows the grammar at the top of \Cref{fig:syntax}.
Nonterminals are formatted in \nt{blue}, terminals are in \term{\kw{red}}, and grammatical elements are in black. We write \{ \nt{a} \} to indicate zero or more occurrences of \nt{a}, and [\nt{b}] to indicate zero or one occurrence of \nt{b}.

The grammar references \emph{expressions} $e$ and entity references $\eref{E}{s}$, whose abstract syntax is given at the bottom of \Cref{fig:syntax}.
The semantics of a policy $c$ is defined by conversion to an expression $e$, written as
$\pcode{toexp}(c)$.
This function conjoins the \nt{Principal}, \nt{Action}, and \nt{Resource} components of $c$, using $\etrue$ is used if the component is just a variable, along with the conditions \nt{Cond}: \kw{when} expressions are used as given, and \kw{unless} expressions are negated.
For example, the conversion of Policy 3 from \Cref{fig:tinytodo-policies} is $\eand{e_s}{(\eor{\ebinop{\kw{in}}{\principal}{\resource.\kw{readers}}}{\ebinop{\kw{in}}{\principal}{\resource.\kw{writers}}})}$, where $e_s$ is the converted scope $\etrue~\kw{\&\&}~$ $\ebinop{\kw{==}}{\action}{\eref{\kw{Action}}{\kw{"GetList"}}}$ $\kw{\&\&}~\etrue$.

\subsection{Expression semantics}
\label{sec:expr-semantics}

We formalize \cedar expression evaluation as a small-step operational semantics with the judgment $\mu, \sigma \vdash e \step e'$, which states that under \emph{entity store} $\mu$ and \emph{authorization request} $\sigma$ the expression $e$ reduces to expression $e'$.
The entity store $\mu$ is a map from entity references $\eref{E}{s}$ to pairs $(v,h)$, where $v$ is a record value and $h$ is the set of the entity's ancestors in the hierarchy.
An authorization request $\sigma$ is a map from variables $x$ to values $v$, where $v$ is an entity reference when $x$ is either \principal, \action, or \resource, and $v$ is a record when $x$ is \context. Stuck states in the semantics correspond to raised errors in the implementation.
% For example, there are no $\mu$, $\sigma$, and $e$ such that $\mu, \sigma \vdash \ebinop{\kw{has}}{i}{f} \step e$; the expression $\ebinop{\kw{has}}{i}{f}$ is stuck because the \code{has} operator is not defined on integers.
% Our implementation raises a run-time type error in this case.

\Cref{fig:semantics} shows a selection of expression evaluation rules.
We do not show any congruence rules as they are straightforward---evaluation is call-by-value and proceeds left to right.
(While \cedar supports neither I/O nor mutable state, evaluation order affects the occurrence of run-time errors.)

The first line of rules shows that projecting an attribute $f$ from an entity reference requires looking up the entity reference in the store and projecting from the corresponding record.
This will fail if the entity is not present in the store or if its record lacks the requested attribute.
Rules for records are similar.
\tighten

The first rule on the second line shows that values are considered equal if and only if they are syntactically identical.
For entity references this amounts to \emph{nominal}, rather than \emph{structural}, equality since $\mu, \sigma \vdash \ebinop{\kw{==}}{\eref{E_1}{s_1}}{\eref{E_2}{s_2}} \step \efalse$ even when $\mu(\eref{E_1}{s_1}) = \mu(\eref{E_2}{s_2})$.
The second rule on that line shows that $\eis{v_1}{E}$ amounts to a dynamic check of whether $v_1$ has the type $E$.
The last rule handles request variable lookup.

The third line shows that membership in the entity hierarchy,
$\ebinop{\kw{in}}{\eref{E_1}{s_1}}{\eref{E_2}{s_2}}$, evaluates to \kw{true} if $\eref{E_2}{s_2}$ is $\eref{E_1}{s_1}$ or is a member of the ancestors of $\eref{E_1}{s_1}$.
Otherwise, the result is \kw{false}.

Per the fourth line, the \kw{in} operator with its RHS as a \emph{set} of entities evaluates to \kw{true} if the LHS is \kw{in} at least one entity in the set (\kw{false} otherwise).
As usual, conditionals short-circuit; note they are not well-defined for non-boolean guards.
Expressions $\eor{e_1}{e_2}$ and $\eand{e_1}{e_2}$ evaluate equivalently to $\kw{if}~e_1~\kw{then}~\etrue$ $\kw{else}~(\econd{e_2}{\etrue}{\efalse})$ and $\kw{if}~e_1~\kw{then}~(\econd{e_2}{\etrue}{\efalse})$ $\kw{else}~\efalse$, respectively.
\tighten

\begin{figure}
    \begin{displaymath}
        \begin{array}{l@{~}c}
            (1)&
            \inferrule
                {\mu(\eref{E}{s}) = (\{ ..., \mapping{f}{v}, ... \},\_)}
                {\mu, \sigma \vdash \ehas{\eref{E}{s}}{f} \step \etrue\\\\
                \mu, \sigma \vdash \eref{E}{s}.f \step v} \qquad

            \inferrule
                {\mu(\eref{E}{s})~\mathrm{undef},~\mathit{or}\\\\
                \mu(\eref{E}{s}) = \{ \mapping{g_1}{v_1}, ..., \mapping{g_n}{v_n} \}\\
                f \not\in \{ g_1, ..., g_n \}}
                {\mu, \sigma \vdash \ehas{\eref{E}{s}}{f} \step \efalse} \\\\

            % (2)&
            % \begin{array}{c}
            % \inferrule
            %     {}
            %     {\mu,\sigma \vdash \{ ..., f:v, ... \}.f \step v} \\[1ex]

            % \inferrule
            %     {}
            %     {\mu,\sigma \vdash \ebinop{\kw{has}}{\{ ..., f:v, ... \}}{f} \step \etrue}
            % \end{array} \qquad

            % \inferrule
            % {v = \{ g_1:v_1, ..., g_n:v_n \}\\\\
            % f \not\in \{ g_1, ..., g_n \} }
            % {\mu, \sigma \vdash \ebinop{\kw{has}}{v}{f} \step \efalse}  \qquad

            (2)&
            \inferrule
                {v_1 = v_2 \Rightarrow v = \etrue\\\\
                v_1 \not= v_2 \Rightarrow v = \efalse}
                {\mu, \sigma \vdash \ebinop{\kw{==}}{v_1}{v_2} \step v} \qquad

            \inferrule
                {(v_1 = \eref{E_1}{s} \wedge E = E_1) \Rightarrow v_2 = \etrue \\\\
                \mathit{(otherwise)} \Rightarrow v_2 = \efalse }
                {\mu, \sigma \vdash \eis{v_1}{E} \step v_2}
            % \qquad
            % \inferrule
            %     {\mu, \sigma \vdash \mathit{matches}(s_1,s_2)}
            %     {\mu, \sigma \vdash \ebinop{\kw{like}}{s_1}{s_2} \step \etrue}
            \qquad
            \inferrule
                {\sigma(x) = v}
                {\mu, \sigma \vdash x \step v}
            \\ \\

            (3)&
            \inferrule
                {\eref{E_2}{s_2} = \eref{E_1}{s_1},~\mathit{or} \\\\
                \mu(\eref{E_1}{s_1}) = (\_,h_1) \\ \eref{E_2}{s_2} \in h_1}
                {\mu, \sigma \vdash \ebinop{\kw{in}}{\eref{E_1}{s_1}}{\eref{E_2}{s_2}} \step \etrue} \qquad

            \inferrule
                {\eref{E_2}{s_2} \not= \eref{E_1}{s_1} \\\\
                \mu(\eref{E_1}{s_1})~\mathrm{undef}\ ~\mathit{or}\ \mu(\eref{E_1}{s_1}) = (\_,h_1)\ \mathit{and}\ \eref{E_2}{s_2} \not\in h_1}
                {\mu, \sigma \vdash \ebinop{\kw{in}}{\eref{E_1}{s_1}}{\eref{E_2}{s_2}} \step \efalse}
            \\ \\

            (4)&
            \begin{array}{c}
            \inferrule
                {(\exists i.\, \mu, \sigma \vdash \ebinop{\kw{in}}{\eref{E}{s}}{\eref{E_i}{s_i}} \step \etrue) \Rightarrow v = \etrue \\\\
                (\forall i.\, \mu, \sigma \vdash \ebinop{\kw{in}}{\eref{E}{s}}{\eref{E_i}{s_i}} \step \efalse) \Rightarrow v = \efalse}
                {\mu, \sigma \vdash \ebinop{\kw{in}}{\eref{E}{s}}{[\eref{E_1}{s_1}, ..., \eref{E_n}{s_n}]} \step v}
            \end{array}
            \quad

            \begin{array}{c}
            \inferrule
                {}
                {\mu, \sigma \vdash \econd{\etrue}{e}{e'} \step e}
                \\[1ex]
                \inferrule
                {}
                {\mu, \sigma \vdash \econd{\efalse}{e'}{e} \step e}
            \end{array}
            % \\ \\

            % (5)&
            % \inferrule
            %     {v \in \{ v_1, ..., v_n \}}
            %     {\mu, \sigma \vdash \ebinop{\kw{contains}}{[v_1, ..., v_n ]}{v} \step \etrue}
            %     \qquad

            % \inferrule
            %     {\{ u_1, ..., u_n \} \subseteq \{ v_1, ..., v_n \}}
            %     {\mu, \sigma \vdash \ebinop{\kw{containsAll}}{[v_1, ..., v_n ]}{[u_1, ..., u_n]} \step \etrue}
            \end{array}
    \end{displaymath}
    \caption{\cedar expression evaluation semantics: Selected rules}
    \label{fig:semantics}
\end{figure}

Boolean operations on sets include \kw{contains}, which checks element membership, \kw{containsAny}, which checks overlap, and \kw{containsAll}, which checks containment.
The \kw{like} operator matches strings, interpreting \kw{*} in the style of the Unix shell.
Integers can be compared, added, subtracted, and multiplied by a constant.

% Operations on sets---two are shown on the last line---are unsurprising.
% Note that \kw{contains} on a set of entity references is \emph{shallow}, just considering direct membership, while \kw{in} on such a set is \emph{deep}, so $\mu, \sigma \vdash \ebinop{\kw{contains}}{v}{\eref{E}{s}} \step \etrue$ implies $\mu, \sigma \vdash \ebinop{\kw{in}}{\eref{E}{s}}{v} \step \etrue$ but not vice versa.

\cedar's semantics is generally forgiving for operations on entity references that do not exist in $\mu$, to minimize the data required to decide a request.
For example, \kw{==} happily compares non-existent references, so operations like \ebinop{\kw{==}}{\action}{\eref{\kw{Action}}{\kw{"foo"}}} can succeed even when $\mu($\eref{\kw{Action}}{\kw{"foo"}}$)$ is undefined.
Evaluation \emph{will} get stuck on a projection $\eref{E}{s}.f$ when $\eref{E}{s}$ does not exist, since it is unclear what value to return.
We considered creating a \emph{null} value, but decided to avoid repeating that ``billion dollar mistake''~\cite{hoare2009null}!

\subsection{Authorization and slicing}
\label{sec:authz-semantics}
\label{sec:policy-slicing}

\cedar's authorization algorithm is given in the following pseudocode:
\[
\begin{array}{l}
${\tiny 1}$ \quad   \pkw{def}~\pcode{authorize}(\mu,C,\sigma) = \\
${\tiny 2}$ \quad    \qquad \pkw{def}~\pcode{evaluate}(e) = v~\pkw{where}~\mu, \sigma \vdash e \step^* v\\
${\tiny 3}$ \quad   \qquad \pkw{let}~C_S = \{ c \mid c \in \pcode{slice}(\sigma,C) \wedge \pcode{evaluate}(\pcode{toexp}(c)) = \etrue \}\\
${\tiny 4}$ \quad   \qquad \pkw{if}~\pcode{forbids}(C_S) = \emptyset \wedge \pcode{permits}(C_S) \not= \emptyset~\pkw{then}~\mathit{Allow}~\pkw{else}~\mathit{Deny}\\
    \end{array}
\]
Procedure \pcode{authorize}() takes policies $C$, a request $\sigma$, and an entity store $\mu$, and returns a decision $\mathit{Allow}$ or $\mathit{Deny}$.
Line~2 defines partial function \pcode{evaluate}$(e)$ to return $v$ if $e$ evaluates to it via the transitive closure of the semantics $\step^*$.
Line~3 uses this function to evaluate the request.
First, it \emph{slices} out the policies in $C$ that are relevant to deciding $\sigma$ (more below).
Then, each policy $c$ in the slice is converted to an expression $e_c$ and evaluated; those policies $c$ which \emph{satisfy} the request (i.e., for which $e_c$ evaluates to $\etrue$) are collected in $C_S$.
Finally, per line~4, the request is allowed if there are no satisfying \kw{forbid} policies and at least one satisfying \kw{permit} policy, else it is denied.\footnote{In the \cedar implementation, the \emph{Allow}/\emph{Deny} decision is coupled with extra diagnostics, which include the IDs of the \emph{determining} policies---the \kw{permit} policies that evaluated to $\etrue$ for \kw{Allow}, or the \kw{forbid} policies that did so for \kw{Deny}---and any policies that exhibited errors when evaluated.}

Rather than evaluate $\sigma$ under every policy $c \in C$, \pcode{authorize} uses subroutine \pcode{slice}$(C,\sigma)$ to quickly select only the policies relevant to $\sigma$.
Recall that the \nt{Principal} and \nt{Resource} parts of the policy scope optionally constrain the \principal{} and \resource, respectively, to be \kw{==} or \kw{in} a particular entity.
Let $\mathit{pof}(c)$ be the entity $\eref{E}{s}$ named in the \nt{Principal} portion of policy $c$, or a special identifier $\kw{Any}$ if no entity is named.
Let $\mathit{rof}(c)$ behave likewise for the \nt{Resource} portion.
Define the key for $c$ to be $\langle \mathit{pof}(c), \mathit{rof}(c) \rangle$.
We can store the policies $C$ as a map $\Sigma$ from keys to sets of policies.
%  (since multiple policies could have the same key).

For a request $\sigma$ we construct a set of keys from the cross-product of the ancestors of $P$ and $R$ in $\mu$, where $P = \sigma(\principal)$ and $R = \sigma(\resource)$.
That is, let $\mu(P) = (\_,h_P)$ and $\mu(R) = (\_,h_R)$, and let $K = (h_P \cup \{P, \kw{Any}\}) \times (h_R \cup \{R, \kw{Any}\})$.
$K$ is the set of keys (each of which is a pair) arising from the cross product of $P$, $R$, and their ancestors or \kw{Any}.
The slice of relevant policies is then $\bigcup_{k \in K} \Sigma(k)$.
Other schemes are possible too, e.g., which take into account the policy conditions.
This scheme performs well when the policy store is large, but $P$ and $R$ have relatively few ancestors.

\subsection{Policy validation}
\label{sec:types-policy}

Users can optionally \emph{validate} their \cedar policies prior to use, to ensure that certain run-time errors will not occur.
Policies are validated against a \emph{schema}, which is a pair $(M, S)$ where $M$ is the \emph{entity schema} and $S$ is the \emph{action schema}.
\begin{itemize}
\item $M$ maps entity types $E$ to pairs $(\tau, P)$ where $\tau$ is the (record) type of $E$'s attributes, and $H$ contains the types of entities that can be ancestors of entities of type $E$.
\item $S$ maps action entities $A$ (of form \eref{\kw{Action}}{s}) to triples $(A_p, A_r, A_x)$ that describe the allowable shape of requests involving $A$.
In particular, $A_p$ is the set of principal entity types that can accompany $A$; $A_r$ is the set of resource entity types that can accompany $A$; and $A_x$ the type of  \context{} records that can accompany $A$.
\end{itemize}
Using the concrete syntax for schemas shown in \Cref{fig:tinytodo-schema}, $M$ is defined by the $\kw{entity}$ declarations, and $S$ is defined by the $\kw{action}$ declarations.

Validating a policy $c$ for a schema $(M,S)$ proceeds in three steps.
First, for each action $A \in \dom{S}$, we produce a set of \emph{request environments} $A_\Gamma$, where request environment $\Gamma \in A_\Gamma$ is a map from variables $x$ to types $\tau$.
$A_\Gamma$ is defined as $\{~\principal{:\,}E_p,\; \resource{:\,}E_r,\; \context{:\,}A_x \mid E_p \in A_p, E_r\in A_r ~\}$.
Second, we convert $c$ into an expression $e = \pcode{toexp}(c)$ (see \Cref{sec:syntax}), and replace all occurrences of variable \action{} in $e$ with $A$; call the result $e_A$.
Finally, we typecheck $e_A$ for each request environment $\Gamma \in A_\Gamma$.
The expression typing judgment has form $\alpha; \Gamma \vdash e : \tau ; \varepsilon$, and is described next.
Typechecking $e_A$ requires that there exists some $\varepsilon$ such that $\emptyset; \Gamma \vdash e_A : \tybool ; \varepsilon$.

\begin{figure}
\begin{subfigure}{\textwidth}
    \begin{displaymath}
        \begin{array}{lc}
            (1) &
            % vars
            % \inferrule
            %     {\Gamma(x) = \tau}
            %     {\alpha; \Gamma \vdash x : \tau; \emptyset}
%            \qquad

            % true and false rules
            \begin{array}{c}
            \inferrule
                {}
                {\alpha; \Gamma \vdash \etrue : \tytrue; \emptyset}
            \\
            \inferrule
                {}
                {\alpha; \Gamma \vdash \efalse : \tyfalse; \varepsilon}
            \\
            \inferrule
            {}
            {\alpha; \Gamma \vdash \ebinop{\kw{==}}{e}{e} : \tytrue; \emptyset}
            \end{array}
            \quad

            % equality rules
            \begin{array}{c}
                \inferrule
                {\alpha; \Gamma \vdash e_1 : E_1; \varepsilon_1\\\\
                 \alpha; \Gamma \vdash e_2 : E_2; \varepsilon_2\\
                 E_1 \neq E_2 }
                {\alpha; \Gamma \vdash \ebinop{\kw{==}}{e_1}{e_2} : \tyfalse; \varepsilon}
            \end{array}
            \quad

            \begin{array}{c}
                \inferrule
            {s_1 \not= s_2}
            {\alpha; \Gamma \vdash \ebinop{\kw{==}}{\eref{E}{s_1}}{\eref{E}{s_2}} : \tyfalse; \varepsilon}
            \end{array}
            \\ \\

            (2) &

            \inferrule
                {\alpha; \Gamma \vdash e_1 : \tau_1; \varepsilon_1\\
                \tau_1 <: \tau \quad \tau_2 <: \tau\\\\
                 \alpha; \Gamma \vdash e_2 : \tau_2; \varepsilon_2\\
                 \text{for some}\;\tau
                 }
                {\alpha; \Gamma \vdash \ebinop{\kw{==}}{e_1}{e_2} : \tybool; \emptyset}
            \qquad

            % is rules
            \inferrule
                {\alpha; \Gamma \vdash e : E_1; \varepsilon\\\\
                (E_1 = E) \Rightarrow \tau = \tytrue \\
                (E_1 \not= E) \Rightarrow \tau = \tyfalse}
                {\alpha; \Gamma \vdash \eis{e}{E} : \tau; \emptyset}
            \\ \\

            (3) &
            % in rules
            \inferrule
                {\alpha; \Gamma \vdash e_1 : E_1; \varepsilon_1\\
                 \alpha; \Gamma \vdash e_2 : E_2; \varepsilon_2\\\\
                 E_1 \neq E_2\\
                 M(E_1) = (\_, H)\\
                 E_2 \notin H}
                {\alpha; \Gamma \vdash \ebinop{\kw{in}}{e_1}{e_2} : \tyfalse; \varepsilon}
            \qquad

            \inferrule
                {\alpha; \Gamma \vdash e_1 : E_1; \varepsilon_1\\
                 \alpha; \Gamma \vdash e_2 : E_2; \varepsilon_2}
                {\alpha; \Gamma \vdash \ebinop{\kw{in}}{e_1}{e_2} : \tybool; \emptyset}
            \\ \\

            (4) &
            % cond
            \inferrule
                {\alpha; \Gamma \vdash e_1 : \tytrue ; \varepsilon_1\\
                 \alpha \cup \varepsilon_1;  \Gamma \vdash e_2 : \tau; \varepsilon_2}
                {\alpha; \Gamma \vdash \econd{e_1}{e_2}{e_3} : \tau; \varepsilon_1 \cup \varepsilon_2}
            \qquad

            \inferrule
                {\alpha; \Gamma \vdash e_1 : \tyfalse ; \varepsilon_1\\
                 \alpha; \Gamma \vdash e_3 : \tau; \varepsilon_3}
                {\alpha; \Gamma \vdash \econd{e_1}{e_2}{e_3} : \tau; \varepsilon_3}
            \\ \\

            (5) &
            \inferrule
                {\alpha; \Gamma \vdash e_1 : \tybool ; \varepsilon_1\\
                 \alpha \cup \varepsilon_1;  \Gamma \vdash e_2 : \tau; \varepsilon_2\\
                 \alpha; \Gamma \vdash e_3 : \tau; \varepsilon_3}
                {\alpha; \Gamma \vdash \econd{e_1}{e_2}{e_3} : \tau, (\varepsilon_1 \cup \varepsilon_2) \cap \varepsilon_3}
            \qquad

            \inferrule
            {\alpha; \Gamma \vdash e : \tau; \varepsilon\\
            \tau <: \tau'}
            {\alpha; \Gamma \vdash e : \tau'; \varepsilon}
            \\ \\

            (6) &
            % has and get-optional for entities,
            \begin{array}{c}
            \inferrule
                {\alpha; \Gamma \vdash e : \tau; \varepsilon \\\\
                \mathit{attribute}(f,\tau) =\, (\mapping{?~f}{\tau_f})}
                {\alpha; \Gamma \vdash \ebinop{has}{e}{f} : \tybool, \{ e.f \}}
            \quad

            \inferrule
                {\alpha; \Gamma \vdash e : \tau; \varepsilon \\\\
                \mathit{attribute}(f,\tau) = (\mapping{\omega f}{\tau_f})\\\\
                 \omega\! = ? \Rightarrow e.f \in \alpha}
                {\alpha; \Gamma \vdash e.f : \tau_f; \emptyset }
            \end{array}
            \;

            \begin{array}{l}
                \mathit{attribute}(f,\tau) = \mapping{\omega f}{\tau_f}\;\text{when} \\
                \quad \tau = \{ ..., \mapping{\omega f}{\tau_f}, ... \},\, \text{or} \\
                \quad \tau = E~\text{and} \\
                \qquad ~~M(E) = (\{ ..., \mapping{\omega f}{\tau_f}, ... \},\_)
            \end{array}
        \end{array}
    \end{displaymath}
    \caption{Typing}
    \label{fig:typing-semantics}
\end{subfigure}
\smallskip
\begin{subfigure}{\textwidth}
    \begin{displaymath}
        \begin{array}{c}
            \inferrule
                {}
                {\tau <: \tau}
            \qquad

            \inferrule
                {\tau <: u\\
                 u <: v}
                {\tau <: v}
            \qquad

            \begin{array}{c}
            \inferrule
                {}
                {\tytrue <: \tybool}
            \\[1ex]
            \inferrule
                {}
                {\tyfalse <: \tybool}
            \end{array}
            \qquad

            \inferrule
              {\tau <: u}
              {\tyset{\tau} <: \tyset{u}}
            \\[3ex]

            \inferrule
              {\tau <: \tau'\\ \omega <: \omega'}
              {\{ \omega_1 f_1  : \tau_1, ... , \omega f : \tau, ..., \omega_n f_n : \tau_n \} <: \{ \omega_1 f_1  : \tau_1, ... , \omega' f : \tau', ..., \omega_n f_n : \tau_n \}}
            \qquad

            \begin{array}{c}
            \inferrule
                {}
                {\omega <: \omega}
            \\[1ex]
            \inferrule
                {}
                {\omega <: ~?}
            \end{array}
        \end{array}
    \end{displaymath}
    \caption{Subtyping}
    \label{fig:subtyping-semantics}
\end{subfigure}
\caption{\cedar expression typing and subtyping: Selected rules}
\label{fig:typing}
\end{figure}

The judgment $\alpha; \Gamma \vdash e : \tau; \varepsilon$ states that under \emph{capability} $\alpha$ and request environment $\Gamma$, expression $e$ has type $\tau$ and produces capability $\varepsilon$ conditioned on $e$ evaluating to $\etrue$.
Selected type rules are given in \Cref{fig:typing-semantics}.
Types are defined at the bottom of \Cref{fig:syntax}, and are notable for the use of \emph{singleton types} \tytrue{} and \tyfalse, and record types with \emph{optional attributes}: a record attribute $\mapping{\omega f}{\tau}$ is optional when $\omega$ is $?$, otherwise it is required; we elide an optionality $\cdot$ when clear from context.
% As discussed below, the type system uses capabilities~\cite{capabilities1999} to ensure safe access to attributes.

% The obvious rules for literals (e.g., integers $i$ have type \tylong), operators (e.g., operator $<$ has type $\tylong \times \tylong \rightarrow \tybool$), and other standard features are elided.
% Typechecking is essentially standard for the majority of \cedar expressions.
% Literals have their obvious types, variables (\principal, \resource, and \context) take their types the request type environment $\Gamma$, and operations such as $>$ or $\kw{like}$ are typechecked in a manner consistent with their runtime types (both arguments are Long for the former and String for the latter).

\paragraph{Boolean singleton types}
Expressions that definitely evaluate to $\etrue$ and $\efalse$ may be given singleton types~\cite{aspinall94singleton} $\tytrue$ and $\tyfalse$, respectively, which are subtypes ($<:$) of $\tybool$.
To make the SMT encoding (\Cref{sec:smt}) tractable, the subtyping rules, shown in \Cref{fig:subtyping-semantics}, support \emph{depth} subtyping but not \emph{width} subtyping, and there is no subtyping among entity types.

For $\kw{==}$, rules on lines (1)-(2) of \Cref{fig:typing-semantics} require that to be type-correct the equated expressions must be \emph{comparable}, in the sense that their types have a shared supertype.
The rules also leverage that two syntactically identical expressions always evaluate to equal values, whereas two entities with different types or identifiers are surely non-equal.

The last rule on line~(2) typechecks \eis{e}{E} with singleton types exclusively, since we know expression $e$'s precise type statically.
Thus \kw{is} is (only) useful when some request environments $\Gamma \in A_\Gamma$ typecheck $e$ such that $E = E_1$, while others do not.
For example, for TinyTodo Policy 1 (\Cref{fig:tinytodo-policies}), the use of \kw{is} is important because \resource\ has type \kw{List} in some request environments and type \kw{Application} in others.

For $\kw{in}$, the first rule on line (3) indicates that $\ebinop{\kw{in}}{e_1}{e_2}$ will have type $\tyfalse$ if the entity type of $e_2$ is not the same as that of $e_1$, and is not among $e_1$'s type's ancestors.
The rules for expressions $\econd{e_1}{e_2}{e_3}$ on line (4) leverage singleton types to ignore unreachable expressions.
The first rule shows that for a $\tytrue$ guard, the conditional will surely evaluate to $e_2$, so $e_3$ can be ignored; the second rule reasons similarly for a $\tyfalse$ guard.%
%\footnote{Rules for $\kw{\&\&}$ and $\kw{||}$ can be precisely derived from the $\kw{if}$ rule.}

Singleton types, and the rules for conditionals in particular, are important for avoiding false alarms when validating a policy.
For example, consider Policy 3 from \Cref{fig:tinytodo-policies}, which converts to the following expression
$$
\begin{array}{l}
\etrue~\kw{\&\&}~(\ebinop{\kw{==}}{\action}{\eref{\kw{Action}}{\kw{"GetList"}}}~\kw{\&\&}~(\etrue~\kw{\&\&}\\
(\eor{\ebinop{\kw{in}}{\principal}{\resource.\kw{readers}}}{\ebinop{\kw{in}}{\principal}{\resource.\kw{writers}}})))
\end{array}
$$
To validate the policy against a schema $(S,M)$ that corresponds to \Cref{fig:tinytodo-schema}, we first consider action \eref{\kw{Action}}{\kw{"CreateList"}}.
We substitute it for $\action$ in the above expression yielding
$$
\etrue~\kw{\&\&}~(\ebinop{\kw{==}}{\eref{\kw{Action}}{\kw{"CreateList"}}}{\eref{\kw{Action}}{\kw{"GetList"}}}~\kw{\&\&}~(...))
$$
We then typecheck this expression under request environment $\Gamma = \mapping{\principal}{\mathit{User}}, \mapping{\resource}{\mathit{Application}}, \mapping{\context}{\{\}}$.
The subexpression $(\ebinop{\kw{==}}{\eref{\kw{Action}}{\kw{"CreateList"}}}{\eref{\kw{Action}}{\kw{"GetList"}}}~\kw{\&\&}~(...))$ is equivalent to
$\econd{\ebinop{\kw{==}}{\eref{\kw{Action}}{\kw{"CreateList"}}}{\eref{\kw{Action}}{\kw{"GetList"}}}}{(...)}{\efalse}$.
When typing this expression, the guard will have type $\tyfalse$ per the first rule on line~(2) of \Cref{fig:typing-semantics}.
So, the second rule on line~(4) ignores the expression $(...)$, written $e_2$ in the rule, giving the whole conditional the type of $\kw{else}$ expression, which is $\tyfalse$.
Doing so is critical:
The $\resource.\kw{readers}$ part of $(...)$ fails to typecheck, since $\Gamma(\resource) = \mathit{Application}$, which has no $\kw{readers}$ attribute.

% Singleton types allow the validator to leverage any correlation between a specific action and the type of $\resource$ (or $\principal$ or $\context$) that may accompany it.
% Consider the other action \eref{\kw{Action}}{\kw{"GetList"}} in $S$.
% The substituted policy expression becomes
% $$
% \etrue~\kw{\&\&}~(\ebinop{\kw{==}}{\eref{\kw{Action}}{\kw{"GetList"}}}{\eref{\kw{Action}}{\kw{"GetList"}}}~\kw{\&\&}~(...))
% $$
% In this case, the equality has type $\tytrue$, so we type subexpression $(...)$ per the first rule on line~(4).
% In this case we are doing it under request environment $\Gamma = \principal: \mathit{User}, \resource: \mathit{List}, \context: \{\}$.
% Since $\Gamma(\resource) = \mathit{List}$, which \emph{does} have a $\kw{readers}$ attribute, all is well.
% \TODO{Somewhere make a comment that this is basically path sensitive typing. Probably we should talk about assumptions on the running time of policy typing.}

\paragraph{Capabilities for accessing optional attributes}
%
% Records and entities may have optional attributes, indicated in their types via \emph{optionality} $\omega$: if $\omega$ is $?$ then the attribute may absent.
The expression $\ehas{e}{f}$ checks whether attribute $f$ is present in a record/entity $e$, and the type system records the effect of this check using \emph{capabilities}~\cite{capabilities1999}.
In judgment $\alpha; \Gamma \vdash e : \tau; \varepsilon$, the $\alpha$ represents the set of optional attributes definitely available when $e$ is evaluated, and $\varepsilon$ is the set of attributes $e$ proves to be available if it evaluates to $\etrue$.

The first rule on line~(6) of \Cref{fig:typing-semantics} types $\ehas{e}{f}$, producing capability $\varepsilon = \{ e.f \}$ when $e$'s type has an optional attribute $f$.%
\footnote{Two other rules, not shown, give this expression type $\tytrue$ when $e$'s type definitely has attribute $f$, and type $\tyfalse$ when it definitely does not, both producing an empty capability $\emptyset$.}
The second rule on line~(6) types expression $e.f$: If $f$ is an optional attribute (i.e., $\omega = ~?$) then $e.f$ must be in the capability $\alpha$.
The way it gets there is via the rules for conditionals on lines~(4)-(5).
The first rules on lines~(4) and~(5) take capability $\varepsilon_1$ proved for the guard $e_1$ and add it to $\alpha$ when typing $e_2$.
This makes sense because $e_2$ is only evaluated when $e_1$ evaluates to $\etrue$.
Putting it all together, when typing expression $\econd{\ebinop{\kw{has}}{e}{f}}{e.f}{\efalse}$ the guard has type $\tybool$ and capability $\{e.f\}$, so the rule on line~(5) types $\kw{then}$-expression $e.f$ under capability $\alpha \cup \{e.f\}$, ensuring the second rule on line~(6) applies.

The rules for conditionals on lines~(4) and~(5) prove the available capability $\varepsilon$ by leveraging the presence of singleton types.
The first rule on line~(4) says that when $e_1$ has type $\tytrue$ we know that $e_2$ will be evaluated, so the capability of the entire conditional is $\varepsilon_1 \cup \varepsilon_2$.
The second rule says that when $e_1$ has type $\tyfalse$ we know that $e_3$ will be evaluated.
The final capability in this case is $\varepsilon_3$ and not $\varepsilon_1 \cup \varepsilon_3$ because attributes in $\varepsilon_1$ are only available if $e_1$ evaluates to $\etrue$, which we know it does not.
The first rule on line~(5) combines the reasoning of these two.
Since the guard has type $\tybool$ we cannot be sure which branch will be evaluated, so we merge the capabilities of $\tytrue$ and $\tyfalse$ cases using intersection, i.e., $(\varepsilon_1 \cup \varepsilon_2) \cap \varepsilon_3$.

In general, per rules on lines~(1)--(3), we can give an expression $e$ of type $\tyfalse$ an arbitrary capability set $\varepsilon$ because doing so is vacuously sound: $\varepsilon$ contains the available attributes \emph{only if $e$ evaluates to $\etrue$}, which we know it does not.
Such rules add very useful precision, especially when handling \kw{\&\&} and \kw{||} expressions.

% The type rules permit testing for multiple attributes together before proceeding to accesses any attributes.
% For example, we can write $\ebinop{\&\&}{(\ebinop{\&\&}{\ebinop{\kw{has}}{e}{f}}{\ebinop{\kw{has}}{e}{g}})}{(\ebinop{||}{e.f}{e.g})}$ to check for the presence of two attributes before the primary condition using those attributes.
% Typechecking this expression requires combining the $\varepsilon$ capability sets from $\ebinop{\kw{has}}{e}{f}$ and $\ebinop{\kw{has}}{e}{g}$ to form the $\alpha$ capability set for $\ebinop{||}{e.f}{e.g}$.
% Intuitively, the capability $\varepsilon$ provided by an expression $\ebinop{\&\&}{e_1}{e_2}$ should be $\varepsilon_1 \cup \varepsilon_2$.
% If we again define $\kw{\&\&}$ in terms of $\kw{if}$, then the capability for the expression is $(\varepsilon_1 \cup \varepsilon_2) \cap \varepsilon_3$ where $\varepsilon_3$ is the capability of $\efalse$.
% Our typing rule for $\efalse$ on line~(1) allows it to have \emph{any} capability, so the intersection of capability sets, and therefore the capability set for $\ebinop{\&\&}{e_1}{e_2}$ may be any subset of $\varepsilon_1 \cup \varepsilon_2$.

\subsection{Properties}
\label{sec:properties}

Since the authorizer is part of an application's security \emph{trusted computing base} (TCB), we take extra steps to confirm its design and implementation are correct.
We formalize \cedar in Lean~\cite{lean4} and use it to prove the following properties:

\textbf{Forbid trumps permit}: If any \kw{forbid} policy evaluates to \etrue, the request is denied.

\textbf{Default deny}: If no \kw{permit} policy evaluates to \etrue, the request is denied.

\textbf{Explicit allow}: If a request is allowed, some \kw{permit} policy evaluated to \etrue.

\textbf{Sound slicing}: $\forall c \in C.\; c \not\in \pcode{slice}(C,\sigma)$ implies $c$ cannot satisfy request $\sigma$.

\textbf{Validation soundness}: Given a policy $c$ and request $\sigma$, let $e_A = \pcode{toexp}(c)[\action \mapsto A]$, where $A = \sigma(\action)$, and let $\Gamma_A$ be a request environment conforming to $\sigma$. If $\emptyset, \Gamma_A \vdash e_A : \tybool, \varepsilon$ under entity type map $M$, then $\mu, \sigma \vdash e_A \step^* v$ for all entity stores $\mu$ conforming to $M$, and for which entity references in $c$ and $\sigma$ are defined.

Carrying out these proofs, and in particular the proof of validation soundness, revealed several bugs.
Moreover, writing down a formal spec forced us to think deeply about corner cases, and the process revealed other subtle bugs in our semantics.
We also use the Lean spec for testing our implementation, written in Rust, as described below in \Cref{sec:full-cedar}.

\subsection{Full \cedar}
\label{sec:full-cedar}

For simplicity, this section has omitted discussion of three of \cedar's features: \emph{templates}, \emph{extension types}, and \emph{action groups}; all three are modeled in the full Lean formalization.

Templates are policies with one or more named \emph{slots}.
A template is similar in spirit to a SQL \emph{prepared statement}, and can be \emph{linked} into a complete \cedar policy by filling each slot with a \cedar value.
At present, \cedar supports only two named slots, \kw{?principal} and \kw{?resource}, which may appear only in the \nt{Principal} and \nt{Resource} parts of the grammar in lieu of specific entities \eref{E}{s}.
This restriction makes it easy to typecheck unlinked templates, and to index template-linked policies.

Extension types provide a uniform mechanism for extending the language.
\cedar currently supports IP addresses and decimal numbers with this mechanism.
Extension-typed values are created with function calls, e.g., \code{ip("1.2.3.4")} and \code{decimal("138.22")}, and operated on with method calls, e.g., \code{resource.amt.lessThan(context.amt)} and \code{context.addr.isInRange(principal.net)}.
The \cedar evaluator implementation provides a plugin mechanism for new extension types.

\cedar actions are entities, so they can also be arranged hierarchically into groups.
Action group memberships are specified in the schema so they can be leveraged during validation.

The Lean model is executable by compilation to native code, so we use extensive differential random testing~\cite{mckeeman1998differential,yang2011finding} to confirm that our Rust code and Lean model agree.
We write input generators (for policies, entity stores, and requests) in the style of \citet{palka11test}, and use the cargo fuzz framework~\cite{cargo-fuzz} to test that equal inputs map to equal outputs.
We also use cargo fuzz to test properties directly on the Rust code, e.g., the ``round trip'' property that a pretty-printed Cedar abstract syntax tree parses back to itself.
This testing regime has been fruitful, uncovering nearly two dozen bugs since the project's inception, and forcing us (via CI) to keep the Lean spec and proofs up to date.

%TODO: Rust implementation details, maybe to include
%         \item The ``ancestors'' relation as the only transitively closed relation (can be indexed for fast evaluation). Can implement other transitively closed relations using attributes + entities.
%         \item Maybe also Big Int semantics for arithmetic ops (with validation identifying potential overflow)

% But what if the application wants a stricter semantics---for example, one that enforces erroring \kw{forbid} policies?
% This is easy to implement by examining the authorization diagnostics and returning \kw{Deny} if a forbid policy errored.
% Notably, if \cedar enforced a stricter semantics, there would be no way to recover the more forgiving semantics from the authorizer output.
% The only solution would be a separate authorization algorithm.
% So \cedar chooses the skip-on-error semantics for uniformity and flexibility.\tighten

\newminted[smtblock]{lisp}{fontsize=\footnotesize{},mathescape,escapeinside=@@}
\newcommand{\sym}[1]{\ensuremath{\hat{#1}}}

\newcommand{\smt}[1]{{\color{gray}\text{\mintinline[fontsize=\small{},mathescape,escapeinside=@@]{lisp}{#1}}}}
\newcommand{\smtFnE}[1]{\ensuremath{#1}\xspace}
\newcommand{\smtFnD}[1]{\ensuremath{#1_D}\xspace}

\newcommand{\mkIdFn}{\ensuremath{\iota}\xspace}
\newcommand{\mkId}[1]{\ensuremath{\mkIdFn(#1)}}

\newcommand{\smtTyFn}{\smtFnE{\mathcal{T}}}
\newcommand{\smtTy}[1]{\ensuremath{\smtTyFn(#1)}}
\newcommand{\declTyFn}{\smtFnD{\mathcal{T}}}
\newcommand{\declTy}[1]{\ensuremath{\declTyFn(#1)}}

\newcommand{\atyMap}{\ensuremath{S}\xspace}
\newcommand{\etyMap}{\ensuremath{M}\xspace}
\newcommand{\tuple}[1]{\ensuremath{\langle #1 \rangle}}
\newcommand{\schema}{\ensuremath{\Sigma}\xspace}
\newcommand{\env}{\ensuremath{\Gamma}\xspace}

\newcommand{\smtVarFn}{\smtFnE{\mathcal{V}}}
\newcommand{\smtVar}[1]{\ensuremath{\smtVarFn(#1)}}
\newcommand{\declVarFn}{\smtFnD{\mathcal{V}}}
\newcommand{\declVar}[1]{\ensuremath{\declVarFn(#1)}}

\newcommand{\smtUFFn}{\smtFnE{\mathcal{F}}}
\newcommand{\smtUF}[1]{\ensuremath{\smtUFFn(#1)}}
\newcommand{\declUFFn}{\smtFnD{\mathcal{F}}}
\newcommand{\declUF}[1]{\ensuremath{\declUFFn(#1)}}

\newcommand{\smtExprFn}{\smtFnE{\mathcal{E}_{\env, \etyMap}}}
\newcommand{\smtExpr}[1]{\ensuremath{\smtExprFn(#1)}}

\newcommand{\ifOkFn}{\ensuremath{\mathit{ifOk}}\xspace}
\newcommand{\ifOk}[2]{\ensuremath{\ifOkFn(#1, #2)}}
\newcommand{\liftedType}[1]{\ensuremath{\mathit{lift}(\mathit{type}(#1))}}
\newcommand{\getFn}{\ensuremath{\mathit{val}}\xspace}
\newcommand{\get}[1]{\ensuremath{\getFn(#1)}}
\newcommand{\isSome}[1]{\ensuremath{\mathit{isSome}(#1)}}
\newcommand{\field}[2]{\ensuremath{\mathit{field}(#1, #2)}}
\newcommand{\subexpr}[1]{\ensuremath{\mathit{subExpr}(#1)}}
\newcommand{\swfFn}{\ensuremath{\mathit{wf}}\xspace}
\newcommand{\swf}[1]{\ensuremath{\swfFn(#1)}}
\newcommand{\acyclic}[1]{\ensuremath{\mathit{wfa}(#1)}}
\newcommand{\transitive}[1]{\ensuremath{\mathit{wft}(#1)}}
\newcommand{\smtImplies}[2]{\ensuremath{\mathit{implies}(#1, #2)}}
\newcommand{\smtAnd}[2]{\ensuremath{\mathit{and}(#1, #2)}}
\newcommand{\reduce}[2]{\ensuremath{\etyMap, \env \vdash #1 \downarrow #2}}

\section{Symbolic Compilation}
\label{sec:smt}

This section presents the \cedar symbolic compiler.
The compiler works in three stages.
First, it encodes \cedar types, entity stores, and requests as SMT types and variables (\Cref{sec:smt:input}).
Next, it uses this mapping to reduce well-typed \cedar expressions to well-typed SMT terms (\Cref{sec:smt:rules}).
Finally, it constrains the SMT representation of the entity hierarchy to be \emph{sufficiently well-formed} (\Cref{sec:smt:swf}), using only ground constraints (i.e., those without quantifiers).
The compiler's encoding is decidable, sound, and complete (\Cref{sec:smt:thms}).
This result, the first of its kind for a non-trivial policy language, is made possible by \cedar's controls on expressiveness, and by leveraging invariants ensured by \cedar's policy validator.

\subsection{Encoding types, entity stores, and requests}\label{sec:smt:input}

\cedar's types are designed to enable a direct translation to SMT types, as shown in \Cref{fig:symbolic-types}.
Primitives and sets are encoded as built-in SMT types (as well as CVC5's theory of sets which is not standardized~\cite{cvc5}), and entities and records as SMT algebraic datatypes.
An entity type becomes a datatype with a string field \code{euid} that represents the entity ID.
A record type becomes a datatype that encodes a required attribute of type $\tau$ as a field of type $\smtTy{\tau}$, and an optional attribute as a field of type $\smt{(Option } \smtTy{\tau} \smt{)}$.
The translation uses the function \mkIdFn to ensure that all occurrences of a given \cedar type map to the same SMT type.\tighten

The symbolic compiler leverages this type mapping to represent the request and entity store as a set of \emph{uninterpreted functions and constants}.
\Cref{fig:symbolic-vars} shows how to generate this representation for a request environment \env and entity schema \etyMap (see \Cref{sec:types-policy}).
The function \declVar{\env} encodes each variable $x \in \env$ as an uninterpreted constant of type \smtTy{\env(x)}.
The function \declUF{\etyMap} encodes the attributes and ancestors of each entity type $E \in \etyMap$ as uninterpreted functions:
\declUF{E, R} maps entities of type \smtTy{E} to their attributes, while \declUF{E, E'} maps entities to their ancestors of type \smtTy{E'}.
Together, these functions and constants represent the set of all possible concrete stores and requests that conform to \etyMap and \env.
We refer to them respectively as the \emph{symbolic store} for \etyMap and \emph{symbolic request} for \env.\tighten

\begin{example}\label{ex:smt-decls}
  Consider the entity schema \etyMap for the TinyTodo schema in \Cref{fig:tinytodo-schema}, and the request environment \env for the action \code{Action::"GetList"}.
  The symbolic store and request for \etyMap and \env are defined as follows, with parts of the encoding omitted for brevity, and names prettified for readability:
  \begin{smtblock}
    (declare-datatype User ((User (eid String))))              ; Entity types
    (declare-datatype Team ((Team (eid String))))
    (declare-datatype List ((List (eid String))))              ; Record types
    (declare-datatype ListRecord ((ListRecord (readers Team) (editors Team) (owner User))))
    (declare-fun listAttrs (List) ListRecord)                  ; Attribute functions
    (declare-fun userInTeam (User) (Set Team))                 ; Ancestor functions
    (declare-fun teamInTeam (Team) (Set Team))
    (declare-const principal User)                             ; principal constant
    (declare-const resource List)                              ; resource constant
  \end{smtblock}
  % \begin{smtblock}
  %   ; ---------- Entity types ----------
  %   (declare-datatype User ((User (eid String))))
  %   (declare-datatype Team ((Team (eid String))))
  %   (declare-datatype List ((List (eid String))))
  %   ; ---------- Record types ----------
  % MWH: We dropped this since it doesn't appear in policies
  %   (declare-datatype UserRecord ((UserRecord (name String))))
  %   (declare-datatype ListRecord ((ListRecord (readers Team) (editors Team) (owner User))))
  %   ; ---------- Attribute functions ----------
  %   (declare-fun userAttrs (User) UserRecord)
  %   (declare-fun listAttrs (List) ListRecord)
  %   ; ---------- Ancestor functions ----------
  %   (declare-fun userInTeam (User) (Set Team))
  %   (declare-fun teamInTeam (Team) (Set Team))
  %   ; ---------- principal and resource constants ----------
  %   (declare-const principal User)
  %   (declare-const resource List)
  % \end{smtblock}
\end{example}

\begin{figure}
  \begin{flalign*}
    \smtTy{\tybool}      & := \smt{Bool}
      \quad\  \smtTy{\tylong}     := \smt{(_ BitVec 64)}
      \quad\ \smtTy{\tystring}    := \smt{String}
      \quad\ \smtTy{\tyset{\tau}} := \smt{(Set } \smtTy{\tau} \smt{)} \\
    \smtTy{E}            & := \mkId{E} \text{ where } E \text{ is an entity type}\\
    \smtTy{R}            & := \mkId{R} \text{ where } R = \{ \omega_1 f_1:\tau_1, \ldots, \omega_n f_n:\tau_n \}\\
    \declTy{E} & := \smt{(declare-datatype } \mkId{E} \smt{ ((} \mkId{E} \smt{ (eid String))))}\\
    \declTy{R} & := \smt{(declare-datatype } \mkId{R} \smt{ ((} \mkId{R} \, \declTy{R, f_1} \,\ldots\, \declTy{R, f_n} \smt{)))}\\
    \declTy{R, f_i} & := \smt{(} \mkId{R, f_i} \ \smtTy{\tau_i} \smt{)}
      \text{ where } R = \{ \ldots, \cdot f_i:\tau_i, \ldots \} \\
    \declTy{R, f_i} & := \smt{(} \mkId{R, f_i} \ \smt{(Option }\smtTy{\tau_i} \smt{))}
      \text{ where } R = \{ \ldots, ? f_i:\tau_i, \ldots \}
  \end{flalign*}
  \caption{Translating a \cedar type to an SMT type. The function \smtTyFn takes as input a \cedar type and returns the corresponding SMT type expression. The function \declTyFn generates SMT type declarations. SMT syntax is rendered in code font, and \mkIdFn maps its arguments to unique SMT identifiers. }
  \label{fig:symbolic-types}
\end{figure}

\begin{figure}
  \begin{minipage}[t]{.8\linewidth}
  \begin{flalign*}
    \declVar{\env} & := \{ \declVar{x, \tau} \mid \env(x) = \tau \} &\\
    \declVar{x, \tau}  & :=  \smt{(declare-const } \mkId{x, \tau} \ \smtTy{\tau} \smt{)} &\\
    \declUF{\etyMap}  & := \bigcup\, \{ \declUF{E, R, P} \mid \etyMap(E) = (R, P) \} &\\
    \declUF{E, R, P}  & :=  \{ \declUF{E, R} \} \cup \{ \declUF{E, E'} \mid E' \in P\} &\\
    \declUF{E, R}     & :=  \smt{(declare-fun } \mkId{E, R} \smt{ (} \smtTy{E}\smt{) } \smtTy{R} \smt{)} &\\
    \declUF{E, E'} & :=  \smt{(declare-fun } \mkId{E, E'} \smt{ (} \smtTy{E}\smt{) } \smtTy{\tyset{E'}} \smt{)} &
  \end{flalign*}
  \end{minipage}%
  \begin{minipage}[t]{.2\linewidth}
  \begin{flalign*}
    \smtVar{x, \tau} & := \mkId{x, \tau} \\
    & \\[1ex]
    \smtUF{E, R} &:= \mkId{E, R} \\
    \smtUF{E, E'} &:= \mkId{E, E'}
  \end{flalign*}
  \end{minipage}
  \caption{Representing entity stores and requests as uninterpreted functions and constants. \declVar{\env} introduces an uninterpreted constant for each variable in the request environment; \declUF{\etyMap} introduces a set of uninterpreted functions for each entity type $E$. $\etyMap(E) = (R,P)$ gives the record and set of ancestor entity types for $E$. }
  \label{fig:symbolic-vars}
\end{figure}

\subsection{Encoding expressions}\label{sec:smt:rules}

The compiler uses the rules in \Cref{fig:symbolic-compilation} to reduce a \cedar expression $e$ to an SMT term \sym{e} under a given symbolic store and request.
This reduction is formalized using a big-step operational semantics: the judgement \reduce{e}{\sym{e}} states that under the symbolic entity store for \etyMap and symbolic request for \env, $e$ reduces to \sym{e}.
The rules assume $e$ is well-typed, and define a total function from well-typed expressions to well-typed terms.
If $e$ has type $\tau$ under \etyMap and \env, then \reduce{e}{\sym{e}} holds, and \sym{e} is a well-typed term of type $\smt{(Option }\smtTy{\tau}\smt{)}$ (\Cref{thm:well-typed-terms}).
We use the \smt{Option} type to encode errors: the term \sym{e} evaluates to \smt{none} on an error not ruled out by the validator (\Cref{sec:properties}).

\begin{theorem}\label{thm:well-typed-terms}
Let $e$, \env, and \etyMap be an expression, environment, and entity schema such that $\alpha; \env \vdash e : \tau, \varepsilon$ for some $\alpha$. Then, there is a well-typed SMT term \sym{e} of type $\smt{(Option }\smtTy{\tau}\smt{)}$ such that \reduce{e}{\sym{e}}.\tighten
\end{theorem}

\Cref{fig:symbolic-compilation} shows a selection of symbolic compilation rules.
Omitted rules are defined straightforwardly in terms of the corresponding SMT operators.
For example, \cedar equality \kw{==} reduces to SMT equality \kw{=} (not shown).\tighten

The rules on line 1 handle variables and entity references, translating them to terms of the form $\smt{(some } \sym{v}\smt{)}$.
A variable $x$ translates to the uninterpreted constant \smtVar{x, \env(x)}, which represents the value of $x$ in the symbolic request.
The entity reference $\eref{E}{s}$ translates to an application of the constructor for the SMT datatype \smtTy{E} to the string $s$.
This constructs a term representing the entity of type $E$ with entity ID $s$ (\Cref{fig:symbolic-types}).\tighten

The rules on line 2 translate attribute operations to terms of the form $\ifOk{\sym{e}}{\sym{v}}$.
The resulting terms encode error propagation by evaluating to \smt{none} if $e$ errors.
For an optional attribute $f$, the expression $e.f$ translates to \sym{f}, which retrieves the optional value of $f$ from the record or entity \get{\sym{e}}.
The expression \ebinop{\kw{has}}{e}{f} translates to a term \sym{b} that evaluates to \smt{true} only if \sym{f} is a \smt{some} value.
Required attributes are handled analogously.\tighten

The rule on line 3 translates the hierarchy membership test $\ebinop{\kw{in}}{e_1}{e_2}$ on entities.
The generated term evaluates to \smt{true} when $\sym{e}_1$ and $\sym{e}_2$ evaluate to entity terms $\sym{v}_1$ and $\sym{v}_2$ satisfying one of two conditions:
$\sym{v}_1$ equals $\sym{v}_2$, or $\sym{v}_1$ has ancestors of type $E_2$ that include $\sym{v}_2$.
This encodes a strongly typed version of the dynamic \kw{in} semantics given on line 4 of \Cref{fig:semantics}.\tighten

Finally, the rules on lines 4 and 5 handle conditional expressions $\econd{e_1}{e_2}{e_3}$.
This follows a standard symbolic semantics with merging~\cite{leanette}.
If $\sym{e}_1$ reduces to \smt{(some true)} or \smt{(some false)}, the conditional translates to $\sym{e}_2$ or $\sym{e}_3$, respectively.
Otherwise, it translates to an \smt{ite} term that selects $\sym{e}_2$ or $\sym{e}_3$ based on the value of $\sym{e}_1$, or errors if $\sym{e}_1$ is \smt{none}.\tighten

\begin{example}\label{ex:smt-term}
  Consider the symbolic store and request from \Cref{ex:smt-decls}, and the expression \code{principal in resource.readers} from Policy 3 in \Cref{fig:tinytodo-schema}.
  The symbolic compiler translates this expression to the term
  \smt{(some (set.member (readers (listAttrs resource)) (userInTeam principal)))}.
\end{example}

\begin{figure}
  \begin{displaymath}
    \begin{array}{lc}
        (1)&
        \inferrule % var
          {\env(x) = \tau}
          {\reduce{x}{\smt{(some } \smtVar{x, \tau} \smt{)}}}

        \qquad
        \inferrule % entity reference
          {\etyMap(E) = (R, \_)}
          {\reduce{\eref{E}{s}}{\smt{(some (} \mkId{E}\ s \smt{))}}}
        \\ \\

        (2)&
        \inferrule % e has f and e.f when f is required
          {\reduce{e}{\sym{e}} \quad \sym{v} = \get{\sym{e}} \\\\
          \field{f}{\sym{v}} = \langle \cdot, \sym{f} \rangle
          }
          {\reduce{\ebinop{\kw{has}}{e}{f}}{\ifOk{\sym{e}}{\smt{(some true)}}} \\\\
          \ \, \reduce{e.f}{\ifOk{\sym{e}}{\smt{(some } \sym{f} \smt{)}}}}

        \quad\quad
        \inferrule % e has f and e.f when f is optional
          {\reduce{e}{\sym{e}} \quad \sym{v} = \get{\sym{e}} \\\\
          \field{f}{\sym{v}} = \langle ?, \sym{f} \rangle  \quad
          \sym{b} = \isSome{\sym{f}}
          }
          {\reduce{\ebinop{\kw{has}}{e}{f}}{\ifOk{\sym{e}}{\smt{(some } \sym{b} \smt{)}}} \\\\
           \reduce{e.f}{\ifOk{\sym{e}}{\sym{f}}}\quad\,}
        \\ \\

        (3)&
        \inferrule % e1 in e2 when both are entities
          {\reduce{e_1}{\sym{e}_1} \quad \sym{v}_1 = \get{\sym{e}_1} \quad \reduce{e_2}{\sym{e}_2} \quad \sym{v}_2 = \get{\sym{e}_2}\\\\
           \liftedType{\sym{v}_1} = E_1 \quad \liftedType{\sym{v}_2} = E_2 \quad \etyMap(E_1) = (\_, P) \\\\
           \sym{b}_1 =
            \text{if } E_1 = E_2
            \text{ then } \smt{(= } \sym{v}_1 \ \sym{v}_2 \smt{)}
            \text{ else } \smt{false} \\\\
           \sym{b}_2 =
            \text{if } E_2 \in P
            \text{ then } \smt{(set.member } \sym{v}_2 \smt{ (} \smtUF{E_1, E_2} \ \sym{v}_1 \smt{))}
            \text{ else } \smt{false}
          }
          {\reduce{\ebinop{\kw{in}}{e_1}{e_2}}{\ifOk{\sym{e}_1}{\ifOk{\sym{e}_2}{\smt{(some (or } \sym{b}_1\ \sym{b}_2 \smt{))}}}}}
        \\ \\

        (4)&
        \inferrule % if true
          {\reduce{e_1}{\smt{(some true)}} \\\\
           \reduce{e_2}{\sym{e}_2}}
          {\reduce{\econd{e_1}{e_2}{e_3}}{\sym{e}_2}}

        \qquad
        \inferrule % if false
          {\reduce{e_1}{\smt{(some false)}} \\\\
           \reduce{e_3}{\sym{e}_3}}
          {\reduce{\econd{e_1}{e_2}{e_3}}{\sym{e}_3}}
        \\ \\

        (5)&
        \inferrule % if symbolic
          {\reduce{e_1}{\sym{e}_1} \quad \sym{e}_1 \neq \smt{(some true)}  \quad \sym{e}_1 \neq \smt{(some false)} \\\\
          \sym{v}_1 = \get{\sym{e}_1} \quad \reduce{e_2}{\sym{e}_2} \quad \reduce{e_3}{\sym{e}_3}}
          {\reduce{\econd{e_1}{e_2}{e_3}}{\ifOk{\sym{e}_1}{\smt{(ite } \sym{v}_1\ \sym{e}_2 \ \sym{e}_3  \smt{)}}}}
      \end{array}
    \end{displaymath}
    \begin{displaymath}\setlength{\arraycolsep}{.5ex}
      \begin{array}{rcl}

        \get{\sym{e}} &:=&
        \text{if } \sym{e} = \smt{(some }\sym{v}\smt{)}
        \text{ then } \sym{v}
        \text{ else } \smt{(val } \sym{e}\smt{)} \\

        \isSome{\sym{e}} &:=&
        \text{if } \sym{e} = \smt{(some }\sym{v}\smt{)}
        \text{ then } \smt{true}
        \text{ else if } \sym{e} = \smt{(as none }\_\smt{)}
        \text{ then } \smt{false}
        \text{ else } \smt{((_ is some) } \sym{e} \smt{)}\\

        \ifOk{\sym{e}_1}{\sym{e}_2} &:=&
        \text{let } \sym{b} = \isSome{\sym{e}_1}, \sym{e}_3 = \smt{(as none } \mathit{type}(\sym{e}_2) \smt{)} \text{ in } \\
        &&
        \text{if } \sym{b} = \smt{true}
        \text{ then } \sym{e}_2
        \text{ else if } \sym{b} = \smt{false}
        \text{ then } \sym{e}_3
        \text{ else } \smt{(ite } \sym{b} \ \sym{e}_2 \ \sym{e}_3 \smt{)} \\

        \field{f}{\sym{v}} &:=&
          \text{if } \liftedType{\sym{v}} = R = \{\ldots, \omega f : \_, \ldots \}
          \text{ then } \langle \omega, \smt{(} \mkId{R, f} \ \sym{v} \smt{)} \rangle  \\
          &&
          \text{if } \liftedType{\sym{v}} = E \text{ and } \etyMap(E) = (R, \_)
          \text{ then } \field{f}{\smt{(} \smtUF{E, R}\ \sym{v} \smt{)}}
      \end{array}
  \end{displaymath}

  \caption{Compiling \cedar expressions to SMT terms: selected rules. We use \liftedType{\sym{v}} to denote the \cedar type that corresponds to the SMT type of the term $\sym{v}$. }
  \label{fig:symbolic-compilation}
\end{figure}

\subsection{Sufficiently well-formed hierarchies}\label{sec:smt:swf}

The rules in \Cref{fig:symbolic-compilation} encode expression semantics for an arbitrary entity store that conforms to the entity schema \etyMap.
But not all such stores are well-formed under \cedar's semantics, which expects the entity hierarchy to be a DAG.
This requires the ancestor functions \smtUF{E, E'} to collectively define an acyclic, transitive relation on entities.
Leaving these functions unconstrained leads to incompleteness, as shown in \Cref{ex:incomplete}.\tighten

\begin{example}\label{ex:incomplete}
  Consider the expression \code{Team::"A" in Team::"B" && Team::"B" in Team::"A"}.
  This expression can never evaluate to true given a well-formed entity hierarchy (a DAG).
  But its encoding with respect to the symbolic store from \Cref{ex:smt-decls} has a model:
  \begin{smtblock}
    (define-fun teamInTeam ((t Team)) (Set Team) ; maps A -> B and B -> A
      (ite (= t (Team "A")) (set.singleton (Team "B")) (set.singleton (Team "A")))
  \end{smtblock}
  This model does not correspond to any well-formed hierarchy because it contains a cycle between \code{Team::"A"} and \code{Team::"B"}.
  As such, it represents a spurious counterexample (a false alarm) to the assertion that the example expression is always false.\tighten
\end{example}

One solution is to encode acyclicity and transitivity constraints on the ancestor functions directly using quantified formulas.
For example, the following formulas assert well-formedness for the function \code{teamInTeam} from \Cref{ex:smt-decls}:
\begin{smtblock}
  (assert ; acyclicity: $\forall t.\, t \not\in \mathtt{teamInTeam}(t)$
    (forall ((t Team)) (not (set.member t (teamInTeam t)))))
  (assert ; transitivity: $\forall t_1, t_2.\, t_2 \in \mathtt{teamInTeam}(t_1) \to \mathtt{teamInTeam}(t_2) \subseteq \mathtt{teamInTeam}(t_1)$
    (forall ((t1 Team) (t2 Team))
      (=> (set.member t2 (teamInTeam t1)) (set.subset (teamInTeam t2) (teamInTeam t1)))))
\end{smtblock}
However, the quantifiers make the encoding undecidable, causing timeouts or unknown results.\tighten

We address this by \emph{grounding} the well-formedness constraints based on the following observation: a \cedar expression $e$ accesses only a finite set of entities from the store.
It is sufficient for the store to be well-formed on just this set.
We can over-approximate this set, called the \emph{footprint} of $e$, by collecting all subexpressions of $e$ with an entity type.
For example, the expression in \Cref{ex:incomplete} has footprint \code{{Team::"A", Team::"B"}}, and the expression in \Cref{ex:smt-term} has footprint \code{{principal, resource, resource.readers}}.
Grounding the constraints relative to the footprint maintains decidability while ensuring completeness (\Cref{sec:smt:thms}).

\Cref{fig:wf} shows the grounding function \swf{e, \env, \etyMap}.
The function emits a transitivity constraint for every pair of footprint expressions of type $E_1$ and $E_2$ that may be transitively connected via an entity of type $E_3$.
It emits an acyclicity constraint for every footprint expression of type $E_1$ that may have an ancestor of the same type.
Together, these two constraints force the relation defined by the ancestor functions to be the transitive closure of an underlying DAG.\tighten

\begin{example}\label{ex:grounding}
  To illustrate grounding, consider again the expression $e$ from \Cref{ex:incomplete}, along with \env and \etyMap from \Cref{ex:smt-decls}.
  The compiler uses $\swf{e, \env, \etyMap}$ to generate the following ground well-formedness assertions:\tighten
  \begin{smtblock}
    (assert (not (set.member (Team "A") (teamInTeam (Team "A")))))
    (assert (not (set.member (Team "B") (teamInTeam (Team "B")))))
    (assert (=> (set.member (Team "B") (teamInTeam (Team "A")))
                (set.subset (teamInTeam (Team "B")) (teamInTeam (Team "A")))))
    (assert (=> (set.member (Team "A") (teamInTeam (Team "B")))
                (set.subset (teamInTeam (Team "A")) (teamInTeam (Team "B")))))
  \end{smtblock}
\end{example}

\newcommand{\andalso}{\textcolor{blue}{\ensuremath{~\wedge~}}}

\begin{figure}
  \begin{displaymath}
    \begin{array}{@{}r@{\ }l}
      \swf{e, \env, \etyMap} :=&
        \text{let } \epsilon =
        \{ \langle \sym{e}_i, E_i, P_i \rangle \mid
           \subexpr{e_i, E_i, e} \andalso
           (\_, P_i) = \etyMap(E_i) \andalso
           \reduce{e_i}{\sym{e}_i} \}
        \text { in } \\
        &
        \{ \acyclic{\sym{e}_1, E_1} \mid
            \langle \sym{e}_1, E_1, P_1 \rangle \in \epsilon \andalso E_1 \in P_1 \} \ \cup \\
        &
        \{\transitive{\sym{e}_1, \sym{e}_2, E_1, E_2, E_3} \mid
          \langle \sym{e}_1, E_1, P_1 \rangle \in \epsilon \andalso
          \langle \sym{e}_2, E_2, P_2 \rangle \in \epsilon \andalso \\
        & \qquad\qquad\qquad\qquad\quad~~
          E_2 \in P_1 \andalso E_3 \in P_1 \cap P_2 \}
        \\
      \acyclic{\sym{e}_1, E_1} :=&
        \text{let } \sym{v}_1 = \get{\sym{e}_1} \text{ in } \\
        &
        \smtImplies
          {\isSome{\sym{e}_1}}
          {\smt{(not (set.member } \sym{v}_1\ \smt{(} \smtUF{E_1, E_1} \ \sym{v}_1 \smt{)))}} \\
      \transitive{\sym{e}_1, \sym{e}_2, E_1, E_2, E_3} :=&
        \text{let }
          \sym{v}_1 = \get{\sym{e}_1},
          \sym{v}_2 = \get{\sym{e}_2},
          \sym{b} = \smtAnd{\isSome{\sym{e}_1}}{\isSome{\sym{e}_2}} \text{ in } \\
        & \mathit{implies}(
            \smtAnd
            {\sym{b}}
            {\smt{(set.member } \sym{v}_2\ \smt{(} \smtUF{E_1, E_2} \ \sym{v}_1 \smt{))}}, \\
        &\qquad\quad
          \smt{(set.subset (} \smtUF{E_2, E_3} \ \sym{v}_2 \smt{) (} \smtUF{E_1, E_3} \ \sym{v}_1 \smt{))})
        \\
      \smtImplies{\sym{b}_1}{\sym{b}_2} :=&
        \text{if } \sym{b}_1 = \smt{true}
        \text{ then } \sym{b}_2
        \text{ else if } \sym{b}_1 = \smt{false}
        \text{ then } \smt{true}
        \text{ else } \smt{(=> } \sym{b}_1 \ \sym{b}_2 \smt{)} \\
      \smtAnd{\sym{b}_1}{\sym{b}_2} :=&
        \text{if } \sym{b}_1 = \smt{true}
        \text{ then } \sym{b}_2
        \text{ else if } \sym{b}_1 = \smt{false}
        \text{ then } \smt{false}
        \text{ else } \smt{(and } \sym{b}_1 \ \sym{b}_2 \smt{)}
    \end{array}
  \end{displaymath}
  \caption{Generating ground well-formedness constraints for $e$, \env, and \etyMap. We use \subexpr{e_i, E_i, e} to denote that $e_i$ is a subexpression of $e$ with type $E_i$, i.e., $\alpha; \env \vdash e_i : E_i, \varepsilon$ for some $\alpha$ and $\varepsilon$.\tighten}
  \label{fig:wf}
\end{figure}

\subsection{Soundness and completeness}\label{sec:smt:thms}

The \cedar symbolic encoding is sound and complete for standard policy analyses, such as equivalence or subsumption~\cite{zelkova2018}.
\Cref{thm:sound-and-complete-equivalence} states the soundness and completeness result for the equivalence analysis.
The proof of the theorem relies on a more general lemma, which we omit for brevity.
The general lemma states that the encoding itself is sound and complete with respect to the concrete semantics of \cedar, following prior work~\cite{leanette}.
This lemma can be used to show that other analyses based on the encoding are also sound and complete.
We have proved soundness and completeness of the encoding on paper.
We have also implemented the symbolic encoder in Lean, and proofs of soundness and completeness using the Lean encoder are underway, as of this writing.\tighten

\begin{theorem}\label{thm:sound-and-complete-equivalence}
  Let $e_1$ and $e_2$ be well-typed boolean expressions under environment \env and entity schema \etyMap.
  Let $\sym{e}_1$ and $\sym{e}_2$ be their encodings where \reduce{e_1}{\sym{e}_1} and \reduce{e_2}{\sym{e}_2}.
  Let $\phi$ be the term $\smt{(= } \mathit{isTrue}(\sym{e}_1)\  \mathit{isTrue}(\sym{e}_2) \smt{)}$, where $\mathit{isTrue}(\sym{e}) := \smt{(= (some true) } \sym{e} \smt{)}$.
  Then, the assertions $\swf{\phi, \env, \etyMap} \cup \{ \smt{(not }\phi\smt{)}\}$ are unsatisfiable iff either $e_1$ and $e_2$ are both true or both untrue on all well-formed inputs conforming to \env and \etyMap.\tighten
\end{theorem}

\section{Evaluation}
\label{sec:eval}

We evaluate \cedar by comparing it against two prominent open-source, general-purpose authorization languages, \fga~\cite{openFGA} and \rego~\cite{opa}, on three example sets of policies.%
\footnote{We report the results of some small experiments comparing \cedar's expressiveness and performance against cloud service providers' purpose-built policy languages in \Cref{sec:related}.}
We briefly discuss how these examples are encoded in each language (\Cref{sec:eval:example-apps}), and we compare the performance in terms of evaluation time for authorization requests (\Cref{subsec:perf-eval}).
We also examine the performance impact of policy slicing (\Cref{sec:eval:policyslicing}) and the performance of our SMT encoding (\Cref{sec:eval:smt}).
% \cdiss{We intend for the reader to also compare readability, but we leave that implicit because it's hard to make strong claims without a user study.}
The code and data used for evaluation is available at \url{https://github.com/cedar-policy/}.\tighten

\subsection{Example applications}
\label{sec:eval:example-apps}
We use the following examples to compare \cedar, \fga, and \rego.
\iftoggle{extended}{
  Full authorization models for each of these examples are given in \Cref{sec:code-appendix}.
}{
  Full authorization models for each of these examples are given in the appendix of an extended version of this paper~\cite{cedar-extended}.
}

\subsubsection{\gdrive example}
\label{sec:eval:gdrive-app}
The \gdrive example~\cite{openFGA-gdrive} comes from \fga; we reimplemented it in \cedar and \rego.
In this application, Users are organized into Groups, and Documents are organized into Folders.
Users may own Documents and Folders, and Users and Groups can be granted view access to Documents and Folders.
View access is transitive for both Folders and Groups: view access to a Folder entails view access to any sub-Folders and contained Documents; and likewise, Users inherit view access from their Group parents.
Other permissions are not transitive: a Folder's owner(s) can create new Documents inside the Folder, but owner(s) of the Folder's parent Folder(s) do not inherit this permission.
Ownership also comes with implied permissions: a Document's owner or owners of the Document's parent Folder(s) can view, share, or write to the Document.\footnote{\fga's provided authorization model does not consider owners of transitive parent Folder(s) for write and share, but we modified the example to consider transitive parents (in both \fga and \cedar), which is in line with the semantics of Google Drive permissions. \Cref{fig:fga-gdrive-model}(a) shows the original \fga declaration for clarity, rather than our modification.}
Finally, Documents may be marked as ``public'' in which case any User has view access.\tighten

\fga encodes these permissions partly in the \emph{authorization model} and partly in the \emph{tuples} representing relationships between specific entities.
This is somewhat analogous to how \cedar has both \emph{policies} and \emph{entity data} respectively.
For example, the \fga declaration of the \code{doc} type in \Cref{fig:fga-gdrive-model}(a) indicates that owners of a document, or owners of the document's parent folder, can write to and share the document.
In \cedar, we express this permission using the policy in \Cref{fig:cedar-gdrive-policy}(b).\tighten

\begin{figure}

% \begin{subfigure}{0.7\textwidth}
% \centering
\begin{tabular}{c|c}
\begin{minipage}{.58\textwidth}
% \begin{subfigure}
\begin{fgablock}
type doc
  relations
  define owner: [user]
  define parent: [folder]
  define viewer: [user, user:*, group#member]
  define can_change_owner: owner
  define can_read: viewer or owner or viewer from parent
  define can_write: owner or owner from parent
  define can_share: owner or owner from parent
\end{fgablock}
% \caption{Declaration of the \code{doc} type in the \fga \gdrive authorization model}
% \end{subfigure}
\end{minipage} &
\begin{minipage}{.37\textwidth}
% \begin{subfigure}
% \centering
% \vspace{1em}
\begin{cedarblock}
permit (
  principal,
  action in [Action::"readDocument",
            Action::"writeDocument",
             Action::"shareDocument"],
  resource)
when {
  resource in
     principal.ownedDocuments ||
  resource in
     principal.ownedFolders };
\end{cedarblock}
% \caption{\cedar policy controlling \code{write} and \code{share} permissions in \gdrive}
% \end{subfigure}
\end{minipage} \\ \\
\parbox{.56\textwidth}{\small\centering (a) \fga \gdrive declaration of the \code{doc} type} &
\parbox{.37\textwidth}{\small\centering (b) \cedar policy controlling \code{write} and \code{share} permissions in \gdrive}
\end{tabular}
\caption{Selected \gdrive permission rules in \fga and \cedar}
\label{fig:fga-gdrive-model}
\label{fig:cedar-gdrive-policy}
\end{figure}

There are many ways to encode the \gdrive authorization model in \cedar; we implemented two.
The first encoding uses static policies over entity data that embody specific permissions, such as a User being granted view access to a Document.
The second one uses \cedar templates to express these permissions; the application would grant view access to a new User or Group by linking a \cedar template.
These encodings represent different tradeoffs in the design space.
We discuss these tradeoffs, including the impact on performance, in \Cref{sec:eval:policyslicing}.\tighten

The encodings of the \gdrive and \github examples in \rego are similar to the one for TinyTodo, discussed below.

\subsubsection{\github example}
\label{sec:github_example}
\fga also provides a \github example~\cite{openFGA-github} which we reimplemented in \cedar and \rego.
In this application, Users and Repositories are organized into Teams and Organizations.
Users and Teams may be granted read/triage/write/maintain/admin privileges on Repositories, and Users or Organizations may be granted read/write/admin privileges on Organizations.
Repositories inherit the permissions of their (Organization) owners.\tighten

As in the \gdrive example, these permissions are encoded partly in policies and partly in entity data.
For \cedar, we make one simplification over \fga's encoding: whereas \fga supports multiple Repository owners of type User or Organization, we require that repositories have a single Organization owner.
As we did for \gdrive, for \github we implemented two different \cedar encodings: one using only static policies, and another using \cedar templates.\tighten

\subsubsection{TinyTodo example}

\begin{figure}
\begin{subfigure}{0.45\textwidth}
\begin{fgablock}
type list
  relations
  define owner: [user]
  define editor: [user,team#member] or owner
  define reader: [user,team#member,user:*] or owner or editor
  define get_list: reader
  define update_list: editor
  define create_task: editor
  define update_task: editor
  define delete_task: editor
  define edit_shares: owner
  define delete_list: owner
\end{fgablock}
\caption{\fga TinyTodo \code{list} type declaration}
\label{fig:fga-tinytodo}
\end{subfigure}
~~\vrule~~
\begin{subfigure}{0.48\textwidth}
\begin{regoblock}
public_actions :=
  ["Action::\"CreateList\"",
   "Action::\"GetLists\""]

allow = true {
    input.Request.Action in public_actions
}

allow = true {
    input.Request.Resource.Owner ==
    input.Request.Principal
}
\end{regoblock}
\caption{\rego TinyTodo policy fragment equivalent to \cedar policies 0 and 1 in \Cref{sec:overview}}
\label{fig:rego-policy0}
\end{subfigure}
\begin{subfigure}{0.75\textwidth}
\vspace{1em}
\begin{regoblock}
allow = true {
    some group in input.Request.Resource.Writers
    group in graph.reachable(input.Data.Groups, [input.Request.Principal])
}
\end{regoblock}
\caption{\rego TinyTodo policy fragment for computing membership transitively}
\label{fig:rego-transitive}
\end{subfigure}
\caption{\fga and \rego code for TinyTodo}
\end{figure}

We also reimplemented the TinyTodo example application (\Cref{sec:overview}) in \fga and \rego.
The \fga translation was straightforward; the most important part is the definition of the \code{list} type, shown in \Cref{fig:fga-tinytodo}.
This defines the hierarchy wherein all owners of a list are also editors and readers, and all editors are also readers.\tighten

Most of the \rego translation was also straightforward, resulting in \rego code like in \Cref{fig:rego-policy0}.
We encoded the graph of team membership using a series of JSON objects with \code{super} and \code{users} attributes.
Unlike \cedar and \fga, \rego does not have any built in notion of a transitive relation.
This can be encoded by either using \rego's standard library graph algorithms, or pre-computing transitive closure over all input data to \rego,
 moving a substantial chunk of authorization logic out of \rego and into the application.
We encode the transitivity within the policy, using graph reachability as in \Cref{fig:rego-transitive}.

\subsection{Performance evaluation}
\label{subsec:perf-eval}

We compare the authorization performance of \cedar, \rego, and \fga on each of the application examples just presented.
We consider the time it takes to authorize a request when all policies and entity data are available in memory.
We thus factor out any time taken to retrieve policies and data from stable storage (if kept there) to focus our comparison on language-level evaluation time.

For each of the application examples, we implemented a random generator that produces entity and request data in the appropriate format, complying with each application's assumed invariants.
Our random generators produce edges in the entity graph (e.g., user-in-group relations, folder-viewable-by-user relations, etc.) in proportion to the number of possible edges in the graph, so the number of edges increases superlinearly as more entities are added.
As we conduct performance testing, we ensure that the responses (allow/deny) returned by each of the authorization engines agree for each request; this confirms that the translation is correct and the respective implementations are solving the same underlying authorization problems.\tighten

\subsubsection{Experimental setup}
\label{sec:eval:experimental-setup}
We conducted performance tests on an Amazon EC2 {\ttfamily\small m5.4xlarge} instance running Amazon Linux 2.
We used \cedar version 3.0.1, \rego version 0.61.0, and \fga commit \code{bbb4a07}.%
\footnote{Based on communication with the \fga developers, the \fga in-memory datastore is intended mainly for debugging and is not optimized, but obviously it is a fairer comparison than would be \fga's persistent-database-backed datastore.}
For each datapoint in the graphs in \Cref{fig:main-perf-results}, we generated 200 separate datastores, and 500 random \code{is_authorized()} requests for each, for a total of 100,000 requests.
We time the execution of the core \code{is_authorized()} operation in \cedar, \fga, and \rego, not including the time required to (for instance) initialize the entity data, parse the \fga authorization model or \cedar/\rego policies, or perform HTTP-related processing.
We report the median and p99 (99th percentile) execution times across these 100,000 requests.\tighten

\subsubsection{Results}
Averaged across all tested input sizes, and considering the median performance, \cedar is \cedarfasterthanfgagdrive, \cedarfasterthanfgagithub, or \cedarfasterthanfgatinytodo faster than \fga (for \gdrive, \github, and TinyTodo respectively).
By the same measure, \cedar is \cedarfasterthanregogdrive, \cedarfasterthanregogithub, or \cedarfasterthanregotinytodo faster than \rego, respectively.

Drilling down further, we also see a marked difference in scaling behavior.
In the \gdrive case, \rego's performance deteriorates from median 76\micros with 5 Users/Groups/Documents/Folders, to median 676\micros with 50 Users/Groups/Documents/Folders.
Similarly, at p99, \rego's performance deteriorates from 391\micros to 1933\micros.
\rego's scaling behavior on \github shows similar trends.
Meanwhile, \fga scales well in the median case, with only a modest increase from 89\micros to 219\micros for \gdrive as the number of entities increases, or 235\micros to 746\micros for \github.
Interestingly, increasing the input size affects \fga's p99 performance more strongly, with the \gdrive p99 increasing from 283\micros to 3012\micros as the number of entities increases.
In contrast to both \rego and \fga, \cedar scales well (with our default, non-templates-based encoding): it handles \gdrive authorization requests in a median 4.0\micros with 5 Users/Groups/Documents/Folders, increasing only to 5.0\micros with 50 Users/Groups/Documents/Folders.
Likewise, its \github performance is around median 11.0\micros across this entire input range.
\cedar's p99 performance is also strong and consistent, under 10\micros (\gdrive) or 20\micros (\github) for all tested input sizes.\tighten

\begin{figure}
\begin{subfigure}{0.31\textwidth}
    \centering
    \includegraphics[width=\textwidth]{results/gdrive/times\_vs\_num\_entities}
    \caption{\gdrive example}
    \label{fig:gdrive-times}
\end{subfigure}
\begin{subfigure}{0.31\textwidth}
    \centering
    \includegraphics[width=\textwidth]{results/github/times\_vs\_num\_entities}
    \caption{\github example}
    \label{fig:github-times}
\end{subfigure}
\begin{subfigure}{0.31\textwidth}
    \centering
    \includegraphics[width=\textwidth]{results/tinytodo/times\_vs\_num\_entities}
    \caption{TinyTodo example}
    \label{fig:tinytodo-times}
\end{subfigure}
\caption{Performance results for \cedar, \fga, and \rego on the \gdrive, \github, and TinyTodo examples}
\label{fig:main-perf-results}
\end{figure}

The \gdrive, \github, and TinyTodo examples rely on transitive relations in the authorization logic.
Since \cedar and \fga provide primitives for reasoning about transitive relationships, they are better equipped to handle these policies.
In \rego, the user has to encode the transitivity of rules within the policy, resulting in both a reduction of clarity and increased running time.
A user of \rego could instead opt to maintain the transitive closure as an invariant on the inputs to \rego, but this comes at the cost of moving a chunk of the authorization logic outside of the authorization policy language.
Doing this improves the scaling behavior of \rego a great deal.
For these three examples, \rego policies are hard to read and do not scale as the other two solutions.\tighten

\subsection{Policy slicing}
\label{sec:eval:policyslicing}

The sound policy slicing scheme presented in \Cref{sec:policy-slicing} does not benefit our \cedar \gdrive and \github examples because their policies do not have scope-level constraints on \principal{} and \resource{}.
However, policy slicing does benefit the template-based encodings of these examples discussed in \Cref{sec:eval:example-apps}.
(We name these variants \gdrivet and \githubt, respectively.)
This is because linked templates have \principal{} or \resource{} constraints in the policy scope, which serve as policy indexing keys.\tighten

\gdrivet contains 4 static policies and 1 template whereas \githubt contains 3 static policies and 5 templates. We use templates to grant specific permissions to individual principals. For instance, we write the following template to grant a certain User, Team, or Organization write access to a repository.

\begin{cedarblock}
  permit (principal in ?principal, action in Action::"writeRepository", resource == ?resource);
\end{cedarblock}

We implement the policy slicing algorithm in \Cref{sec:policy-slicing} and evaluate its performance by randomly generating template links and requests.
For each possible pair of principal and resource entities, we generate a template link for that pair with probability 0.05.
So, the expected number of template-linked policies is quadratic in the number of entities per entity type.
We use the same experimental setup as described in \Cref{sec:eval:experimental-setup}.\tighten

We evaluate policy slicing by comparing our default \cedar encodings with no templates (``\cedar''), our template-based \cedar encodings without policy slicing (``\cedar (Templates)''), and our template-based \cedar encodings with policy slicing enabled (``\cedar (Policy Slicing)'').
The results are shown in \Cref{fig:eval:policyslicing}.
With templates but without policy slicing, \cedar's performance is much poorer: there are many more policies to evaluate, and this more than offsets the performance benefit from the template encodings' slightly less complex policies and slightly less entity data.
With policy slicing enabled, most of the policies are excluded from the slice, and overall performance returns to comparable to our default \cedar encoding without templates---averaged across all tested input sizes, the \cedar (Policy Slicing) median is about 18\% faster than the \cedar median for \github, but about 94\% slower than \cedar for \gdrive.\tighten
% \shaobo{if we think about it, the probability that a policy is kept is inversely quadratic to the number of entities per type, and the number of policies is quadratic, making the expected number constant.}
% \shaobo{do we need to show the time to build policy indices?}

  \begin{figure}
    \begin{subfigure}{0.48\textwidth}
        \centering
        \includegraphics[width=\textwidth]{results/gdrive-slicing/times\_vs\_num\_entities}
        \caption{\gdrivet example}
        \label{fig:eval:gdrive-slicing}
    \end{subfigure}
    \begin{subfigure}{0.48\textwidth}
        \centering
        \includegraphics[width=\textwidth]{results/github-slicing/times\_vs\_num\_entities}
        \caption{\githubt example}
        \label{fig:eval:github-slicing}
    \end{subfigure}
    \caption{Performance results of sound policy slicing}
    \label{fig:eval:policyslicing}
    \end{figure}

\subsection{SMT encoding}
\label{sec:eval:smt}

The \cedar symbolic compiler can be used to implement many different SMT-based analyses.
One particularly useful analysis is checking that refactorings are done correctly.
This requires analyzing the whole policy set, and is thus a good way to evaluate the scalability of our encoding.
To perform this evaluation, we consider a refactoring operation on each example application.\tighten

We illustrate the refactoring using the \github model discussed in~\Cref{sec:github_example}.
Suppose we started with the two policies in \Cref{fig:cedar-original-policies} and wanted to replace them with the policy in \Cref{fig:cedar-refactored-policy}.
This refactoring is correct: it doesn't change the set of requests that satisfy the policy set under the given schema.
The refactoring may seem obvious, but it is only correct because none of the actions listed have children in the schema.
If we modify the schema to add a child action, the refactoring is no longer valid, because of the semantics of \code{in} on sets (see line 4 in \autoref{fig:semantics}):\tighten

\begin{schemablock}
  action "bug_inducing" in ["writeRepository"]
    appliesTo { principal: [User], resource: [Repository] }
\end{schemablock}

We perform this same refactoring (expanding policies with lists of actions into multiple policies comparing actions using \emph{==}) to all of the policies in each of our applications.
The overall timing results and times spent in the SMT solver are given in \Cref{tbl:smt_runtimes}.

\begin{figure}
\begin{subfigure}[t]{0.5\textwidth}
\begin{cedarblock}
permit (
  principal,
  action == Action::"writeRepository",
  resource)
when { principal in resource.owner.writers };
permit (
  principal,
  action == Action::"triageRepository",
  resource)
when { principal in resource.owner.writers };
\end{cedarblock}
\caption{Original policies}
\label{fig:cedar-original-policies}
\end{subfigure}
~\vrule~
\begin{subfigure}[t]{0.45\textwidth}
\begin{cedarblock}
  permit (
    principal,
    action in
      [Action::"writeRepository",
      Action::"triageRepository"],
    resource)
  when {
    principal in resource.owner.writers
  };
\end{cedarblock}
\caption{Refactored policy}
\label{fig:cedar-refactored-policy}
\end{subfigure}
\caption{Analysis example: Equivalence after a refactoring of \github policies}
\end{figure}

\begin{table}[h!]
\caption{Median running times (50 trials) of SMT encoding + solving}
\label{tbl:smt_runtimes}

% \begin{tabular}{ | c | c | c | c | }
%     \hline
%     SMT     & Github     & GDrive    & TinyTodo      \\
%     valid   &  31.7(2.7) & 26.7(0.6) & 62.0(5.0) \\
%     invalid &  37.8(2.7) & 32.7(0.9) & 69.4(6.2) \\
%     \hline
% \end{tabular}
% \quad
% \begin{tabular}{ | c | c | c | c | }
%     \hline
%     Comb.   & Github     & GDrive    & TinyTodo    \\
%     valid   &  58.4(7.4) & 41.7(1.0) & 111.8(11.1) \\
%     invalid &  66.3(7.4) & 50.4(1.4) & 121.8(13.9) \\
%     \hline
% \end{tabular}

\begin{tabular}{ c | c | c | c }
  Refactoring & \github (ms) & \gdrive (ms) & TinyTodo (ms) \\ \hline
  Correct     &  27.1+33.8  & 16.4+28.9   & 52.2+67.4     \\
  Buggy       &  29.9+40.9  & 19.6+34.9   & 56.2+77.3     \\
\end{tabular}
\end{table}

All experiments are run 50 times on a Intel(R) Xeon(R) Platinum 8124M CPU @ 3.00GHz.
We use CVC5~\cite{cvc5} version 1.0.5 as our solver.
Note that the invalid refactorings generally take longer to solve because we invoke the solver once for each action and the invalid cases have the additional action, \code{bug_inducing}.\tighten

\section{Related work}
\label{sec:related}

\newcommand*\emptycirc{\tikz\draw (0,0) circle (0.7ex);}
\newcommand*\halfcirc{\tikz\draw[fill] (0,0)-- (90:0.7ex) arc (90:270:0.7ex) -- cycle ;}
\newcommand*\fullcirc{\tikz\fill (0,0) circle (0.7ex);}

\newcommand{\zero}{-}
\newcommand{\one}{\ensuremath{\emptycirc}}
\newcommand{\two}{\ensuremath{\halfcirc}}
\newcommand{\three}{\ensuremath{\fullcirc}}

Authorization policy languages have been studied extensively over the last few decades.
\Cref{tab:related-comparison} presents a qualitative comparison of \cedar to closely related prior languages.
The comparison is along six axes:
(1) \emph{Expressiveness} considers support for common access control models, \code{forbid} policies, and search and string matching operations;
(2) \emph{Syntax} considers the readability of the language syntax;
(3) \emph{Performance} considers support for policy indexing and the running time of a policy in terms of policy and input size, both typical and worst cases;
(4) \emph{Formal} considers whether the language has a formal semantics and metatheory for security-relevant properties;
(5) \emph{Validation} considers support for policy validation, ranging from linting to sound type checking; and
(6) \emph{Analysis} considers whether the language supports semantic policy analysis.
Elements of this characterization are necessarily subjective, and complete information was not always available.\tighten

\cedar natively supports RBAC and ABAC, can encode ReBAC, and supports \code{forbid} policies.
\cedar has rudimentary string matching, and limits searching to membership in sets and the entity hierarchy.
\cedar's syntax directly expresses authorization concepts in simple terms.
\cedar policies generally have linear-time performance; set containment operations can make running times quadratic in the worst case.
\cedar has a formal semantics, with metatheory proved in Lean.
It offers a sound schema-based policy validator, as well as a sound and complete encoding for the full language to a decidable fragment of SMT-LIB.\tighten

\begin{table}
  \caption{Qualitative comparison with prior authorization languages}
  \label{tab:related-comparison}
  \begin{tiny}
  \noindent
  \begin{tabular}{l|c|c|c|c|c|c|c|c|l|c|c|c}
  Language & \multicolumn{6}{c|}{Expressiveness} & Syntax & \multicolumn{2}{c|}{Performance} & Formal & Validation & Analysis \\
  & RBAC & ABAC & ReBAC & \texttt{forbid} & search & str & & index & eval wc (typ) && \\ \hline
  \cedar & \three & \three & \two & \three & \one & \one & \three & yes & poly (lin) & yes & \three & \three \\
  XACML & \two & \three & \zero & \three & \two & \three & \one & partial & poly & partial & $\pm$\two & \one \\
  OPA/Rego & \two & \three & \two & \two & \three & \two & \two & partial & exp (poly) & no & \three & \zero \\
  Zanzibar & \three & \zero & \three & \three & \zero & \zero & \three & yes & lin & no & \three & $\pm$\two \\
  Ponder & \three & \three & \three & \three & \three & \two & \two & yes & exp (poly) & no & \two & $\pm$\one \\
  % AWS IAM & \zero & \three & \zero & \three & \one & \one & \two & no & poly (lin) & partial & \one & \two \\
  Cassandra & \three & \two & \two & \zero & \three & $e$ & \two & no & exp (poly) & yes & \three & $\pm$\one \\
  SecureUML & \three & \three & \two & \zero & \one & \two & \one & yes & poly & partial & $\pm$\three & $\pm$\one \\
  \end{tabular}
  ~\\

  \begin{tabular}{ll} \\
    \three & full, native support \\
    \two & partial support, or full support with encoding \\
    \one & minimal support \\
  \end{tabular}%
  \begin{tabular}{ll} \\
    \zero & no support \\
    $e$ & support via extension \\
    $\pm$ & possible support, none now, and unlikely to be full support
    \end{tabular}
\end{tiny}
\end{table}

\textbf{XACML}~\cite{xacml} is an open source declarative policy language.
XACML provides native support for ABAC and \code{forbid} (called \emph{deny}) rules; RBAC can be encoded, but it is unclear how to encode ReBAC.
XACML provides slightly more general search and string-matching combinators compared to \cedar.
XACML's syntax is XML-based, which can be hard to read; ALFA~\cite{alfa} defines a more readable syntax for XACML.\@
XACML policies are labeled with an explicit set of \emph{target} conditions that form the basis of policy indexing, analogously to \cedar policies' scope.
XACML has typed operators, but no schema-style validator (that we could find).
Several prior works developed analyses for XACML policies, leading to partial formalizations of the language~\cite{xacml-formalism}.
\citet{margrave2005} developed Margrave, which uses multi-terminal binary decision diagrams to provide sound satisfiability and change impact analysis for a limited subset of XACML.
\citet{hughes2008automated} analyze a different subset of XACML by translation to SAT.
Their encoding is neither sound nor complete.\tighten

\textbf{Open Policy Agent} (OPA)~\cite{opa} is an open source authorization system that uses a Datalog-based language called \rego.
\rego, like other Datalog-based languages (some are discussed below), is more expressive than \cedar, allowing users to define their own notions of evidence prioritization and combination, and data hierarchy.
\rego naturally supports \code{permit}-style ABAC policies, and is powerful enough to encode RBAC, ReBAC, and \code{forbid} policies.
\rego supports expressive string matching and search via recursive rules.
\rego's added expressiveness is a blessing but also a curse.
It provides no built-in authorization concepts such as `allow' or `deny', leaving policy authors to make their own choices to encode these concepts, which later policy readers may find difficult to understand (especially if they are not experts).
\rego's evaluation performance can also be hard to anticipate.
While typical uses for authorization should be polynomial~\cite{madden2022datalogauth}, worst-case performance is exponential~\cite{dantsin2001complexity}.
Datalog's resolution algorithm is behind the scenes, so small policy differences can potentially lead to performance surprises~\cite{tekle2010precise,opaissueloop}.
The \rego documentation suggests policy writers limit themselves to a ``linear fragment,'' though it is not precise about what that fragment is~\cite{opalinear}.
OPA supports indexing but rules must adhere to certain restrictions~\cite{sandall2017ruleindex}.
Rego provides a policy validator that leverages JSON schema~\cite{rego-schema}, but it is neither specified carefully nor proved sound.
Producing an analyzer would be difficult, as Datalog program equivalence is undecidable~\cite{SHMUELI1993231}.\tighten

\textbf{Zanzibar}~\cite{zanzibar2019} is a highly scalable, relationship-based access control (ReBAC) system that Google uses to manage permissions for its cloud products.
The system defines authorization in terms of relationships between users and resources.
Zanzibar's implementation is proprietary, but there are several open source clones, such as Ory Keto~\cite{oryketo}, AuthZed SpiceDB~\cite{spicedb}, and Auth0 Fine Grained Authorization (FGA)~\cite{openFGA}.
Zanzibar supports RBAC and ReBAC with a natural syntax, and with policy validation.
It does not support ABAC, string matching, or search.
\code{forbid}-style policies are supported by expressing one relationship in terms of non-membership in some other relation.
Zanzibar can load authorization models lazily, providing a kind of indexing~\cite{zanzibar2019}.
Our experiments show that OpenFGA's performance is roughly linear in the number of tuples involved in relevant relationships.
Zanzibar provides no deep policy analysis now, but it might be possible, e.g., by using the SMT theory of relations~\cite{cvc-relations}.\tighten

\textbf{Ponder}~\cite{10.1007/3-540-44569-2_2} is a declarative object-oriented policy specification language for writing access control policies, with a readable syntax.
It has direct support for ABAC, RBAC, ReBAC, and \code{forbid} policies.
Ponder policy constraints are specified in a subset of the Object Constraint Language~\cite{OCL}, which is typed, has no side effects, supports string matching, and provides a variety of higher-order operators that enable sophisticated searches.
Ponder deployments support resource-based policy indexes~\cite{dulay2001policy}, and while Ponder's powerful operators could result in exponential worst-case performance, typical policies should be polynomial.
Ponder has no formal semantics, but offers some degree of typechecking.
We are not aware of automated analyses for Ponder policies, but there have been analyses developed for OCL via encoding to formal logic for use by SMT solvers~\cite{clavel2010checking,dania2016ocl2msfol}.
These encodings leave out some OCL features, and are otherwise incomplete due to the use of quantifiers.\tighten

\textbf{SecureUML}~\cite{lodderstedt2002secureuml} specifies access rules via UML, in an approach called \emph{model driven security}~\cite{basin2011decade}.
RBAC-style rules can be refined by conditions over attributes, which are specified in OCL\@.
Rules are attached to UML actions, supporting a kind of indexing, and OCL conditions generally encourage polynomial-time performance.
Analysis for SecureUML leverages analysis for OCL constraints, which has the limitations discussed above.

\textbf{Cassandra} is a Datalog-based authorization language parametrized by a choice of \emph{constraint domain} to support arithmetic operations, set operations, and others.
Cassandra can express a variety of access control models and support powerful search and string matching operations, but at the cost of exponential worst-case performance and even greater challenges in building analysis tools.
Cassandra focuses on \code{permit} policies; it is unclear how \code{forbid} policies could be supported.
Cassandra is a \emph{trust management}~\cite{policymaker} system, which handles \emph{decentralized} authorization by potentially consulting (or receiving signed certificates from) multiple policy stores instead of a single, centralized policy store.
Other similar works (also based on Datalog) include \textbf{RT$x$}~\cite{RT}, the Query Certificate Manager (\textbf{QCM})~\cite{qcm}, and Secure Dynamically Distributed Datalog (\textbf{SD3})~\cite{sd3}.
It would be interesting to explore extensions to \cedar to support the decentralized approach.\tighten

\bigskip

\textbf{Cloud service authorization policy languages} are purpose-built to protect resources managed by cloud services, including those from Google Cloud~\cite{gcpiam}, Microsoft Azure~\cite{azurepolicy}, and Amazon Web Services (AWS)~\cite{awsiam}.
\cedar is primarily different in that it is general-purpose: policies can be customized to meet an application's needs via \cedar's user-defined principal, resource, and action types; custom entity hierarchies; and arbitrary entity relationships.
Nevertheless, \cedar has features that are similar to those in cloud-service authorization languages, and supports some of the same patterns.
For example, AWS IAM and Azure policies are also structured around principals/actions/resources, and they support service-specific attributes and roles.
While a full translator is outside of the scope of this work, as a simple exercise we closely examined a subset of the AWS IAM policies in the benchmark collected by \citet{eiers2022quacky} (specifically, the policies under the ``samples/iam'' folder in their repository~\cite{quacky2022github}).
We found that we could hand-translate each of these policies to \cedar fairly directly.
Authorization times on the translated policies were, averaged over 300 runs with 950 entities, 14\micros with random queries and 15\micros with well-formed queries.
These times are in line with those of our experiments in \Cref{subsec:perf-eval}.

\cedar's symbolic compiler (described in \Cref{sec:smt}) likewise bears similarities to previous work on symbolic analysis of cloud-based authorization languages.
\citet{zelkova2018} present the first symbolic analysis of AWS IAM policies, with the aim of detecting policy misconfigurations.
Eiers et al.\@ perform symbolic analysis of AWS IAM and Microsoft Azure policies, using model counting to quantify permissiveness~\cite{eiers2022quantifying}, compare policies' permissiveness~\cite{eiers2022quacky}, or perform policy repair~\cite{eiers2023quantitative}.
Our symbolic compiler for \cedar differs from these in that its encoding is both sound and complete, which is made possible by \cedar's careful language design.
This encoding can be a building block for follow-on analyses; for instance, it would be interesting to build analogues of the policy-misconfiguration or model-counting analyses from these previous works on top of our \cedar encoding.
% Moreover, in this work we describe only the encoding to SMT, and not any additional analyses built on top of this encoding, such as analyses comparable in purpose to the ones in \cite{eiers2022quantifying} and \cite{zelkova2018}.

% Previous text explaining AWS IAM's notations in the table
% The syntax of AWS IAM is based on JSON, making it more readable than XACML but less so than \cedar.
% AWS IAM policies are less expressive than \cedar in many ways, e.g., due to the lack of support for RBAC or ReBAC.
% Performance bounds on policy evaluation are good, though AWS IAM policies are not indexed.
% AWS IAM provides a policy linter but no sound validator.
% Zelkova~\cite{zelkova2018} is a closed-source tool for analyzing properties of AWS IAM policies via translation to SMT formulae; it powers services such as the AWS IAM Access Analyzer~\cite{AA2020}.
% The Zelkova encoding is decidable and sound, but not complete.\tighten

\section{Conclusion}
\label{sec:conclusion}

This paper has presented \cedar, a new authorization policy language whose design carefully balances expressiveness, safety, performance, and analyzability.
\cedar is used at scale in products and services from Amazon Web Services, and is open source.
All code, documentation, and example applications can be found at \url{https://github.com/cedar-policy/}.

\paragraph*{Acknowledgments}

The authors would like to thank the anonymous referees for their helpful feedback on a draft of this paper; Sarah Cecchetti, Julian Lovelock, Abhishek Panday, and Mark Stalzer for their contributions to the development of Cedar; Bhakti Shah for contributions to Cedar's Lean development; and Andrew Gacek for his help in constructing our TinyTodo sample application.

\bibliography{bib}


\begin{thebibliography}{55}

%%% the Web address.

\ifx \showCODEN    \undefined \def \showCODEN     #1{\unskip}     \fi
\ifx \showDOI      \undefined \def \showDOI       #1{#1}\fi
\ifx \showISBNx    \undefined \def \showISBNx     #1{\unskip}     \fi
\ifx \showISBNxiii \undefined \def \showISBNxiii  #1{\unskip}     \fi
\ifx \showISSN     \undefined \def \showISSN      #1{\unskip}     \fi
\ifx \showLCCN     \undefined \def \showLCCN      #1{\unskip}     \fi
\ifx \shownote     \undefined \def \shownote      #1{#1}          \fi
\ifx \showarticletitle \undefined \def \showarticletitle #1{#1}   \fi
\ifx \showURL      \undefined \def \showURL       {\relax}        \fi
% invisible to TeX
\providecommand\bibfield[2]{#2}
\providecommand\bibinfo[2]{#2}
\providecommand\natexlab[1]{#1}
\providecommand\showeprint[2][]{arXiv:#2}

\bibitem[Aspinall(1994)]%
        {aspinall94singleton}
\bibfield{author}{\bibinfo{person}{David Aspinall}.}
  \bibinfo{year}{1994}\natexlab{}.
\newblock \showarticletitle{Subtyping with singleton types}. In
  \bibinfo{booktitle}{\emph{CSL: International Workshop on Computer Science
  Logic}}. \bibinfo{pages}{1--15}.
\newblock


\bibitem[authzed-spicedb(2024)]%
        {spicedb}
authzed-spicedb \bibinfo{year}{2024}\natexlab{}.
\newblock \bibinfo{title}{spicedb}.
\newblock \bibinfo{howpublished}{\url{https://github.com/authzed/spicedb}}.
\newblock
\newblock
\shownote{Open Source, Google Zanzibar-inspired permissions database to enable
  fine-grained access control for customer applications}.


\bibitem[aws-iam(2024)]%
        {awsiam}
aws-iam \bibinfo{year}{2024}\natexlab{}.
\newblock \bibinfo{title}{Access Management -- AWS Identity and Access
  Management (IAM)}.
\newblock \bibinfo{howpublished}{\url{https://aws.amazon.com/iam/}}.
\newblock


\bibitem[azure-policy(2024)]%
        {azurepolicy}
azure-policy \bibinfo{year}{2024}\natexlab{}.
\newblock \bibinfo{title}{Azure policy documentation}.
\newblock
  \bibinfo{howpublished}{\url{https://learn.microsoft.com/en-us/azure/governance/policy/}}.
\newblock


\bibitem[Backes et~al\mbox{.}(2018)]%
        {zelkova2018}
\bibfield{author}{\bibinfo{person}{John Backes}, \bibinfo{person}{Pauline
  Bolignano}, \bibinfo{person}{Byron Cook}, \bibinfo{person}{Catherine Dodge},
  \bibinfo{person}{Andrew Gacek}, \bibinfo{person}{Kasper Luckow},
  \bibinfo{person}{Neha Rungta}, \bibinfo{person}{Oksana Tkachuk}, {and}
  \bibinfo{person}{Carsten Varming}.} \bibinfo{year}{2018}\natexlab{}.
\newblock \showarticletitle{Semantic-based Automated Reasoning for AWS Access
  Policies using SMT}. In \bibinfo{booktitle}{\emph{2018 Formal Methods in
  Computer Aided Design (FMCAD)}}. \bibinfo{pages}{1--9}.
\newblock
\urldef\tempurl%
\url{https://doi.org/10.23919/FMCAD.2018.8602994}
\showDOI{\tempurl}


\bibitem[Barbosa et~al\mbox{.}(2022)]%
        {cvc5}
\bibfield{author}{\bibinfo{person}{Haniel Barbosa}, \bibinfo{person}{Clark~W.
  Barrett}, \bibinfo{person}{Martin Brain}, \bibinfo{person}{Gereon Kremer},
  \bibinfo{person}{Hanna Lachnitt}, \bibinfo{person}{Makai Mann},
  \bibinfo{person}{Abdalrhman Mohamed}, \bibinfo{person}{Mudathir Mohamed},
  \bibinfo{person}{Aina Niemetz}, \bibinfo{person}{Andres N{\"{o}}tzli},
  \bibinfo{person}{Alex Ozdemir}, \bibinfo{person}{Mathias Preiner},
  \bibinfo{person}{Andrew Reynolds}, \bibinfo{person}{Ying Sheng},
  \bibinfo{person}{Cesare Tinelli}, {and} \bibinfo{person}{Yoni Zohar}.}
  \bibinfo{year}{2022}\natexlab{}.
\newblock \showarticletitle{cvc5: {A} Versatile and Industrial-Strength {SMT}
  Solver}. In \bibinfo{booktitle}{\emph{Tools and Algorithms for the
  Construction and Analysis of Systems - 28th International Conference, {TACAS}
  2022, Held as Part of the European Joint Conferences on Theory and Practice
  of Software, {ETAPS} 2022, Munich, Germany, April 2-7, 2022, Proceedings,
  Part {I}}} \emph{(\bibinfo{series}{Lecture Notes in Computer Science},
  Vol.~\bibinfo{volume}{13243})}, \bibfield{editor}{\bibinfo{person}{Dana
  Fisman} {and} \bibinfo{person}{Grigore Rosu}} (Eds.).
  \bibinfo{publisher}{Springer}, \bibinfo{pages}{415--442}.
\newblock
\urldef\tempurl%
\url{https://doi.org/10.1007/978-3-030-99524-9\_24}
\showDOI{\tempurl}


\bibitem[Basin et~al\mbox{.}(2011)]%
        {basin2011decade}
\bibfield{author}{\bibinfo{person}{David Basin}, \bibinfo{person}{Manuel
  Clavel}, {and} \bibinfo{person}{Marina Egea}.}
  \bibinfo{year}{2011}\natexlab{}.
\newblock \showarticletitle{A decade of model-driven security}. In
  \bibinfo{booktitle}{\emph{Proceedings of the 16th ACM symposium on Access
  control models and technologies (SACMAT)}}.
\newblock


\bibitem[Blaze et~al\mbox{.}(1998)]%
        {policymaker}
\bibfield{author}{\bibinfo{person}{Matt Blaze}, \bibinfo{person}{Joan
  Feigenbaum}, {and} \bibinfo{person}{Martin Strauss}.}
  \bibinfo{year}{1998}\natexlab{}.
\newblock \showarticletitle{Compliance checking in the PolicyMaker trust
  management system}. In \bibinfo{booktitle}{\emph{Financial Cryptography}}.
\newblock


\bibitem[cargo-fuzz(2023)]%
        {cargo-fuzz}
cargo-fuzz \bibinfo{year}{2023}\natexlab{}.
\newblock \bibinfo{title}{Rust Fuzz Book}.
\newblock
  \bibinfo{howpublished}{\url{https://rust-fuzz.github.io/book/cargo-fuzz.html}}.
\newblock


\bibitem[Caserio et~al\mbox{.}(2022)]%
        {xacml-formalism}
\bibfield{author}{\bibinfo{person}{Carmine Caserio}, \bibinfo{person}{Francesca
  Lonetti}, {and} \bibinfo{person}{Eda Marchetti}.}
  \bibinfo{year}{2022}\natexlab{}.
\newblock \showarticletitle{A Formal Validation Approach for {XACML 3.0} Access
  Control Policy}.
\newblock \bibinfo{journal}{\emph{Sensors}} \bibinfo{volume}{22},
  \bibinfo{number}{8} (\bibinfo{year}{2022}).
\newblock
\showISSN{1424-8220}
\urldef\tempurl%
\url{https://www.mdpi.com/1424-8220/22/8/2984}
\showURL{%
\tempurl}


\bibitem[Clavel et~al\mbox{.}(2010)]%
        {clavel2010checking}
\bibfield{author}{\bibinfo{person}{Manuel Clavel}, \bibinfo{person}{Marina
  Egea}, {and} \bibinfo{person}{Miguel Angel~Garc{\'\i}a de Dios}.}
  \bibinfo{year}{2010}\natexlab{}.
\newblock \showarticletitle{Checking unsatisfiability for {OCL} constraints}.
\newblock \bibinfo{journal}{\emph{Electronic Communications of the EASST}}
  \bibinfo{volume}{24} (\bibinfo{year}{2010}).
\newblock


\bibitem[Crary et~al\mbox{.}(1999)]%
        {capabilities1999}
\bibfield{author}{\bibinfo{person}{Karl Crary}, \bibinfo{person}{David Walker},
  {and} \bibinfo{person}{Greg Morrisett}.} \bibinfo{year}{1999}\natexlab{}.
\newblock \showarticletitle{Typed Memory Management in a Calculus of
  Capabilities}. In \bibinfo{booktitle}{\emph{Proceedings of the 26th ACM
  SIGPLAN-SIGACT Symposium on Principles of Programming Languages}} (San
  Antonio, Texas, USA) \emph{(\bibinfo{series}{POPL '99})}.
  \bibinfo{publisher}{Association for Computing Machinery},
  \bibinfo{address}{New York, NY, USA}, \bibinfo{pages}{262–275}.
\newblock
\showISBNx{1581130953}
\urldef\tempurl%
\url{https://doi.org/10.1145/292540.292564}
\showDOI{\tempurl}


\bibitem[Cutler et~al\mbox{.}(2024)]%
        {cedar-oopsla}
\bibfield{author}{\bibinfo{person}{Joseph~W. Cutler}, \bibinfo{person}{Craig
  Disselkoen}, \bibinfo{person}{Aaron Eline}, \bibinfo{person}{Shaobo He},
  \bibinfo{person}{Kyle Headley}, \bibinfo{person}{Michael Hicks},
  \bibinfo{person}{Kesha Hietala}, \bibinfo{person}{Eleftherios Ioannidis},
  \bibinfo{person}{John Kastner}, \bibinfo{person}{Anwar Mamat},
  \bibinfo{person}{Darin McAdams}, \bibinfo{person}{Matt McCutchen},
  \bibinfo{person}{Neha Rungta}, \bibinfo{person}{Emina Torlak}, {and}
  \bibinfo{person}{Andrew~M. Wells}.} \bibinfo{year}{2024}\natexlab{}.
\newblock \showarticletitle{{Cedar}: A New Language for Expressive, Fast, Safe,
  and Analyzable Authorization}.
\newblock \bibinfo{journal}{\emph{Proc. {ACM} Program. Lang.}}
  \bibinfo{volume}{8}, \bibinfo{number}{{OOPSLA1}} (\bibinfo{year}{2024}).
\newblock


\bibitem[Damianou et~al\mbox{.}(2001)]%
        {10.1007/3-540-44569-2_2}
\bibfield{author}{\bibinfo{person}{Nicodemos Damianou},
  \bibinfo{person}{Naranker Dulay}, \bibinfo{person}{Emil Lupu}, {and}
  \bibinfo{person}{Morris Sloman}.} \bibinfo{year}{2001}\natexlab{}.
\newblock \showarticletitle{The Ponder Policy Specification Language}. In
  \bibinfo{booktitle}{\emph{Policies for Distributed Systems and Networks}},
  \bibfield{editor}{\bibinfo{person}{Morris Sloman}, \bibinfo{person}{Emil~C.
  Lupu}, {and} \bibinfo{person}{Jorge Lobo}} (Eds.).
  \bibinfo{publisher}{Springer Berlin Heidelberg}, \bibinfo{address}{Berlin,
  Heidelberg}, \bibinfo{pages}{18--38}.
\newblock


\bibitem[Dania and Clavel(2016)]%
        {dania2016ocl2msfol}
\bibfield{author}{\bibinfo{person}{Carolina Dania} {and}
  \bibinfo{person}{Manuel Clavel}.} \bibinfo{year}{2016}\natexlab{}.
\newblock \showarticletitle{{OCL2MSFOL}: a mapping to many-sorted first-order
  logic for efficiently checking the satisfiability of OCL constraints}. In
  \bibinfo{booktitle}{\emph{Proceedings of the ACM/IEEE 19th International
  Conference on Model Driven Engineering Languages and Systems}}.
  \bibinfo{pages}{65--75}.
\newblock


\bibitem[Dantsin et~al\mbox{.}(2001)]%
        {dantsin2001complexity}
\bibfield{author}{\bibinfo{person}{Evgeny Dantsin}, \bibinfo{person}{Thomas
  Eiter}, \bibinfo{person}{Georg Gottlob}, {and} \bibinfo{person}{Andrei
  Voronkov}.} \bibinfo{year}{2001}\natexlab{}.
\newblock \showarticletitle{Complexity and expressive power of logic
  programming}.
\newblock \bibinfo{journal}{\emph{ACM Computing Surveys (CSUR)}}
  \bibinfo{volume}{33}, \bibinfo{number}{3} (\bibinfo{year}{2001}),
  \bibinfo{pages}{374--425}.
\newblock


\bibitem[Dulay et~al\mbox{.}(2001)]%
        {dulay2001policy}
\bibfield{author}{\bibinfo{person}{Naranker Dulay}, \bibinfo{person}{Emil
  Lupu}, \bibinfo{person}{Morris Sloman}, {and} \bibinfo{person}{Nicodemos
  Damianou}.} \bibinfo{year}{2001}\natexlab{}.
\newblock \showarticletitle{A policy deployment model for the {Ponder}
  language}. In \bibinfo{booktitle}{\emph{2001 IEEE/IFIP International
  Symposium on Integrated Network Management Proceedings. Integrated Network
  Management VII. Integrated Management Strategies for the New Millennium (Cat.
  No. 01EX470)}}. IEEE, \bibinfo{pages}{529--543}.
\newblock


\bibitem[Eiers et~al\mbox{.}(2023a)]%
        {eiers2023quantitative}
\bibfield{author}{\bibinfo{person}{William Eiers}, \bibinfo{person}{Ganesh
  Sankaran}, {and} \bibinfo{person}{Tevfik Bultan}.}
  \bibinfo{year}{2023}\natexlab{a}.
\newblock \showarticletitle{Quantitative Policy Repair for Access Control on
  the Cloud}.
\newblock  (\bibinfo{year}{2023}).
\newblock


\bibitem[Eiers et~al\mbox{.}(2022)]%
        {eiers2022quantifying}
\bibfield{author}{\bibinfo{person}{William Eiers}, \bibinfo{person}{Ganesh
  Sankaran}, \bibinfo{person}{Albert Li}, \bibinfo{person}{Emily O'Mahony},
  \bibinfo{person}{Benjamin Prince}, {and} \bibinfo{person}{Tevfik Bultan}.}
  \bibinfo{year}{2022}\natexlab{}.
\newblock \showarticletitle{Quantifying permissiveness of access control
  policies}. In \bibinfo{booktitle}{\emph{Proceedings of the 44th International
  Conference on Software Engineering}}. \bibinfo{pages}{1805--1817}.
\newblock


\bibitem[Eiers et~al\mbox{.}(2023b)]%
        {eiers2022quacky}
\bibfield{author}{\bibinfo{person}{William Eiers}, \bibinfo{person}{Ganesh
  Sankaran}, \bibinfo{person}{Albert Li}, \bibinfo{person}{Emily O'Mahony},
  \bibinfo{person}{Benjamin Prince}, {and} \bibinfo{person}{Tevfik Bultan}.}
  \bibinfo{year}{2023}\natexlab{b}.
\newblock \showarticletitle{Quacky: Quantitative Access Control Permissivness
  Analyzer}. In \bibinfo{booktitle}{\emph{Proceedings of the 37th IEEE/ACM
  International Conference on Automated Software Engineering}} (<conf-loc>,
  <city>Rochester</city>, <state>MI</state>, <country>USA</country>,
  </conf-loc>) \emph{(\bibinfo{series}{ASE '22})}.
  \bibinfo{publisher}{Association for Computing Machinery},
  \bibinfo{address}{New York, NY, USA}, Article \bibinfo{articleno}{163},
  \bibinfo{numpages}{5}~pages.
\newblock
\showISBNx{9781450394758}
\urldef\tempurl%
\url{https://doi.org/10.1145/3551349.3559530}
\showDOI{\tempurl}


\bibitem[Ferraiolo and Kuhn(1992)]%
        {rbac}
\bibfield{author}{\bibinfo{person}{David~F. Ferraiolo} {and}
  \bibinfo{person}{D.~Richard Kuhn}.} \bibinfo{year}{1992}\natexlab{}.
\newblock \showarticletitle{Role-Based Access Control}. In
  \bibinfo{booktitle}{\emph{15th National Computer Security Conference}}.
\newblock


\bibitem[Fisler et~al\mbox{.}(2005)]%
        {margrave2005}
\bibfield{author}{\bibinfo{person}{K. Fisler}, \bibinfo{person}{S.
  Krishnamurthi}, \bibinfo{person}{L.A. Meyerovich}, {and}
  \bibinfo{person}{M.C. Tschantz}.} \bibinfo{year}{2005}\natexlab{}.
\newblock \showarticletitle{Verification and change-impact analysis of
  access-control policies}. In \bibinfo{booktitle}{\emph{Proceedings. 27th
  International Conference on Software Engineering, 2005. ICSE 2005.}}
  \bibinfo{pages}{196--205}.
\newblock
\urldef\tempurl%
\url{https://doi.org/10.1109/ICSE.2005.1553562}
\showDOI{\tempurl}


\bibitem[gcp-iam(2024)]%
        {gcpiam}
gcp-iam \bibinfo{year}{2024}\natexlab{}.
\newblock \bibinfo{title}{Identity and Access Management | IAM | Google Cloud}.
\newblock \bibinfo{howpublished}{\url{https://cloud.google.com/iam/}}.
\newblock


\bibitem[Gunter and Jim(2000)]%
        {qcm}
\bibfield{author}{\bibinfo{person}{{Carl A.} Gunter} {and}
  \bibinfo{person}{Trevor Jim}.} \bibinfo{year}{2000}\natexlab{}.
\newblock \showarticletitle{Policy-directed certificate retrieval}.
\newblock \bibinfo{journal}{\emph{Software - Practice and Experience}}
  \bibinfo{volume}{30}, \bibinfo{number}{15} (\bibinfo{date}{Dec.}
  \bibinfo{year}{2000}).
\newblock


\bibitem[Hoare(2009)]%
        {hoare2009null}
\bibfield{author}{\bibinfo{person}{C.A.R. Hoare}.}
  \bibinfo{year}{2009}\natexlab{}.
\newblock \bibinfo{title}{Null References: The Billion Dollar Mistake}.
\newblock \bibinfo{howpublished}{Presentation at the {QCon} conference}.
\newblock
\urldef\tempurl%
\url{https://www.infoq.com/presentations/Null-References-The-Billion-Dollar-Mistake-Tony-Hoare/}
\showURL{%
\tempurl}


\bibitem[Hu et~al\mbox{.}(2015)]%
        {abac}
\bibfield{author}{\bibinfo{person}{Vincent~C. Hu}, \bibinfo{person}{D.~Richard
  Kuhn}, \bibinfo{person}{David~F. Ferraiolo}, {and} \bibinfo{person}{Jeffrey
  Voas}.} \bibinfo{year}{2015}\natexlab{}.
\newblock \showarticletitle{Attribute-Based Access Control}.
\newblock \bibinfo{journal}{\emph{Computer}} \bibinfo{volume}{48},
  \bibinfo{number}{2} (\bibinfo{year}{2015}), \bibinfo{pages}{85--88}.
\newblock
\urldef\tempurl%
\url{https://doi.org/10.1109/MC.2015.33}
\showDOI{\tempurl}


\bibitem[Hughes and Bultan(2008)]%
        {hughes2008automated}
\bibfield{author}{\bibinfo{person}{Graham Hughes} {and} \bibinfo{person}{Tevfik
  Bultan}.} \bibinfo{year}{2008}\natexlab{}.
\newblock \showarticletitle{Automated verification of access control policies
  using a SAT solver}.
\newblock \bibinfo{journal}{\emph{International journal on software tools for
  technology transfer}} \bibinfo{volume}{10}, \bibinfo{number}{6}
  (\bibinfo{year}{2008}), \bibinfo{pages}{503--520}.
\newblock


\bibitem[Jim(2001)]%
        {sd3}
\bibfield{author}{\bibinfo{person}{Trevor Jim}.}
  \bibinfo{year}{2001}\natexlab{}.
\newblock \showarticletitle{{SD3}: A Trust Management System with Certified
  Evaluation}. In \bibinfo{booktitle}{\emph{IEEE Symposium on Security and
  Privacy}}.
\newblock


\bibitem[Li et~al\mbox{.}(2005)]%
        {RT}
\bibfield{author}{\bibinfo{person}{Ninghui Li}, \bibinfo{person}{John~C.
  Mitchell}, {and} \bibinfo{person}{William~H. Winsborough}.}
  \bibinfo{year}{2005}\natexlab{}.
\newblock \showarticletitle{Beyond proof-of-compliance: security analysis in
  trust management}.
\newblock \bibinfo{journal}{\emph{J. ACM}}  \bibinfo{volume}{52}
  (\bibinfo{year}{2005}).
\newblock


\bibitem[Lodderstedt et~al\mbox{.}(2002)]%
        {lodderstedt2002secureuml}
\bibfield{author}{\bibinfo{person}{Torsten Lodderstedt}, \bibinfo{person}{David
  Basin}, {and} \bibinfo{person}{J{\"u}rgen Doser}.}
  \bibinfo{year}{2002}\natexlab{}.
\newblock \showarticletitle{{SecureUML}: A {UML}-based modeling language for
  model-driven security}. In \bibinfo{booktitle}{\emph{International Conference
  on the Unified Modeling Language}}. \bibinfo{pages}{426--441}.
\newblock


\bibitem[Madden(2022)]%
        {madden2022datalogauth}
\bibfield{author}{\bibinfo{person}{Neil Madden}.}
  \bibinfo{year}{2022}\natexlab{}.
\newblock \bibinfo{title}{Is Datalog a good language for authorization?}
\newblock
  \bibinfo{howpublished}{\url{https://neilmadden.blog/2022/02/19/is-datalog-a-good-language-for-authorization/}}.
\newblock


\bibitem[McKeeman(1998)]%
        {mckeeman1998differential}
\bibfield{author}{\bibinfo{person}{William~M McKeeman}.}
  \bibinfo{year}{1998}\natexlab{}.
\newblock \showarticletitle{Differential testing for software}.
\newblock \bibinfo{journal}{\emph{Digital Technical Journal}}
  \bibinfo{volume}{10}, \bibinfo{number}{1} (\bibinfo{year}{1998}),
  \bibinfo{pages}{100--107}.
\newblock


\bibitem[Meng et~al\mbox{.}(2017)]%
        {cvc-relations}
\bibfield{author}{\bibinfo{person}{Baoluo Meng}, \bibinfo{person}{Andrew
  Reynolds}, \bibinfo{person}{Cesare Tinelli}, {and} \bibinfo{person}{Clark
  Barrett}.} \bibinfo{year}{2017}\natexlab{}.
\newblock \showarticletitle{Relational Constraint Solving in SMT}. In
  \bibinfo{booktitle}{\emph{Automated Deduction -- CADE 26}},
  \bibfield{editor}{\bibinfo{person}{Leonardo de~Moura}} (Ed.).
  \bibinfo{publisher}{Springer International Publishing},
  \bibinfo{address}{Cham}, \bibinfo{pages}{148--165}.
\newblock
\showISBNx{978-3-319-63046-5}


\bibitem[{MITRE}(2023)]%
        {cwe2023top25}
\bibfield{author}{\bibinfo{person}{{MITRE}}.} \bibinfo{year}{2023}\natexlab{}.
\newblock \bibinfo{title}{{CWE} Top 25 Most Dangerous Software Weaknesses}.
\newblock
  \bibinfo{howpublished}{\url{https://cwe.mitre.org/top25/archive/2023/2023_top25_list.html}}.
\newblock


\bibitem[Moura and Ullrich(2021)]%
        {lean4}
\bibfield{author}{\bibinfo{person}{Leonardo~de Moura} {and}
  \bibinfo{person}{Sebastian Ullrich}.} \bibinfo{year}{2021}\natexlab{}.
\newblock \showarticletitle{The Lean 4 Theorem Prover and Programming
  Language}. In \bibinfo{booktitle}{\emph{Automated Deduction -- CADE 28}},
  \bibfield{editor}{\bibinfo{person}{Andr{\'e} Platzer} {and}
  \bibinfo{person}{Geoff Sutcliffe}} (Eds.). \bibinfo{publisher}{Springer
  International Publishing}, \bibinfo{address}{Cham},
  \bibinfo{pages}{625--635}.
\newblock
\showISBNx{978-3-030-79876-5}


\bibitem[OPA(2023)]%
        {opa}
OPA \bibinfo{year}{2023}\natexlab{}.
\newblock \bibinfo{title}{Policy-based control for cloud native environments}.
\newblock \bibinfo{howpublished}{\url{https://www.openpolicyagent.org/}}.
\newblock


\bibitem[OPA-linear(2023)]%
        {opalinear}
OPA-linear \bibinfo{year}{2023}\natexlab{}.
\newblock \bibinfo{title}{Open Policy Agent documentation: Linear fragment}.
\newblock
  \bibinfo{howpublished}{\url{https://www.openpolicyagent.org/docs/latest/policy-performance/linear-fragment}}.
\newblock


\bibitem[OpenFGA(2023a)]%
        {openFGA-github}
OpenFGA \bibinfo{year}{2023}\natexlab{a}.
\newblock \bibinfo{title}{{OpenFGA} {GitHub} sample store}.
\newblock
  \bibinfo{howpublished}{\url{https://github.com/openfga/sample-stores/tree/main/stores/github}}.
\newblock


\bibitem[OpenFGA(2023b)]%
        {openFGA-gdrive}
OpenFGA \bibinfo{year}{2023}\natexlab{b}.
\newblock \bibinfo{title}{{OpenFGA} {G}oogle {D}rive sample store}.
\newblock
  \bibinfo{howpublished}{\url{https://github.com/openfga/sample-stores/tree/main/stores/gdrive}}.
\newblock


\bibitem[OpenFGA(2023c)]%
        {openFGA}
OpenFGA \bibinfo{year}{2023}\natexlab{c}.
\newblock \bibinfo{title}{{OpenFGA}: Relationship-based access control made
  fast, scalable, and easy to use}.
\newblock \bibinfo{howpublished}{\url{https://openfga.dev/}}.
\newblock


\bibitem[ory-keto(2024)]%
        {oryketo}
ory-keto \bibinfo{year}{2024}\natexlab{}.
\newblock \bibinfo{title}{keto}.
\newblock \bibinfo{howpublished}{\url{https://github.com/ory/keto}}.
\newblock
\newblock
\shownote{Open Source (Go) implementation of "Zanzibar: Google's Consistent,
  Global Authorization System"}.


\bibitem[Pa\l{}ka et~al\mbox{.}(2011)]%
        {palka11test}
\bibfield{author}{\bibinfo{person}{Micha\l{}~H. Pa\l{}ka},
  \bibinfo{person}{Koen Claessen}, \bibinfo{person}{Alejandro Russo}, {and}
  \bibinfo{person}{John Hughes}.} \bibinfo{year}{2011}\natexlab{}.
\newblock \showarticletitle{Testing an Optimising Compiler by Generating Random
  Lambda Terms}. In \bibinfo{booktitle}{\emph{Proceedings of the 6th
  International Workshop on Automation of Software Test}}.
\newblock


\bibitem[Pang et~al\mbox{.}(2019)]%
        {zanzibar2019}
\bibfield{author}{\bibinfo{person}{Ruoming Pang}, \bibinfo{person}{Ramon
  Caceres}, \bibinfo{person}{Mike Burrows}, \bibinfo{person}{Zhifeng Chen},
  \bibinfo{person}{Pratik Dave}, \bibinfo{person}{Nathan Germer},
  \bibinfo{person}{Alexander Golynski}, \bibinfo{person}{Kevin Graney},
  \bibinfo{person}{Nina Kang}, \bibinfo{person}{Lea Kissner},
  \bibinfo{person}{Jeffrey~L. Korn}, \bibinfo{person}{Abhishek Parmar},
  \bibinfo{person}{Christina~D. Richards}, {and} \bibinfo{person}{Mengzhi
  Wang}.} \bibinfo{year}{2019}\natexlab{}.
\newblock \showarticletitle{Zanzibar: {G}oogle’s Consistent, Global
  Authorization System}. In \bibinfo{booktitle}{\emph{2019 {USENIX} Annual
  Technical Conference ({USENIX} {ATC} '19)}}. \bibinfo{address}{Renton, WA}.
\newblock


\bibitem[Porncharoenwase et~al\mbox{.}(2022)]%
        {leanette}
\bibfield{author}{\bibinfo{person}{Sorawee Porncharoenwase},
  \bibinfo{person}{Luke Nelson}, \bibinfo{person}{Xi Wang}, {and}
  \bibinfo{person}{Emina Torlak}.} \bibinfo{year}{2022}\natexlab{}.
\newblock \showarticletitle{A Formal Foundation for Symbolic Evaluation with
  Merging}.
\newblock \bibinfo{journal}{\emph{Proc. ACM Program. Lang.}}
  \bibinfo{volume}{6}, \bibinfo{number}{POPL}, Article \bibinfo{articleno}{47}
  (\bibinfo{date}{jan} \bibinfo{year}{2022}), \bibinfo{numpages}{28}~pages.
\newblock
\urldef\tempurl%
\url{https://doi.org/10.1145/3498709}
\showDOI{\tempurl}


\bibitem[Quacky(2022)]%
        {quacky2022github}
Quacky \bibinfo{year}{2022}\natexlab{}.
\newblock \bibinfo{title}{Quacky}.
\newblock
  \bibinfo{howpublished}{\url{https://github.com/vlab-cs-ucsb/quacky/tree/master/samples}}.
\newblock


\bibitem[Rego Policy Language: Schema(2023)]%
        {rego-schema}
Rego Policy Language: Schema \bibinfo{year}{2023}\natexlab{}.
\newblock \bibinfo{title}{Policy Language: Schema}.
\newblock
  \bibinfo{howpublished}{\url{https://www.openpolicyagent.org/docs/latest/policy-language/\#schema}}.
\newblock


\bibitem[Sandall(2017)]%
        {sandall2017ruleindex}
\bibfield{author}{\bibinfo{person}{Torin Sandall}.}
  \bibinfo{year}{2017}\natexlab{}.
\newblock \bibinfo{title}{Optimizing {OPA}: Rule indexing}.
\newblock
  \bibinfo{howpublished}{\url{https://blog.openpolicyagent.org/optimizing-opa-rule-indexing-59f03f17caf3}}.
\newblock


\bibitem[Sandall(2020)]%
        {opaissueloop}
\bibfield{author}{\bibinfo{person}{Torin Sandall}.}
  \bibinfo{year}{2020}\natexlab{}.
\newblock \bibinfo{title}{[OPA issue] Implement loop-invariant code motion
  optimization}.
\newblock
  \bibinfo{howpublished}{\url{https://github.com/open-policy-agent/opa/issues/2094}}.
\newblock


\bibitem[Shmueli(1993)]%
        {SHMUELI1993231}
\bibfield{author}{\bibinfo{person}{Oded Shmueli}.}
  \bibinfo{year}{1993}\natexlab{}.
\newblock \showarticletitle{Equivalence of Datalog queries is undecidable}.
\newblock \bibinfo{journal}{\emph{The Journal of Logic Programming}}
  \bibinfo{volume}{15}, \bibinfo{number}{3} (\bibinfo{year}{1993}),
  \bibinfo{pages}{231--241}.
\newblock
\showISSN{0743-1066}
\urldef\tempurl%
\url{https://doi.org/10.1016/0743-1066(93)90040-N}
\showDOI{\tempurl}


\bibitem[Standard(2013)]%
        {xacml}
\bibfield{author}{\bibinfo{person}{OASIS Standard}.}
  \bibinfo{year}{2013}\natexlab{}.
\newblock \bibinfo{title}{Extensible Access Control Markup Language ({XACML})
  version 3.0}.
\newblock
  \bibinfo{howpublished}{\url{http://docs.oasis-open.org/xacml/3.0/xacml-3.0-core-spec-os-en.html}}.
\newblock


\bibitem[Standard(2014)]%
        {OCL}
\bibfield{author}{\bibinfo{person}{OMG Standard}.}
  \bibinfo{year}{2014}\natexlab{}.
\newblock \bibinfo{title}{Object Constraint Language (OCL), version 2.4}.
\newblock \bibinfo{howpublished}{\url{http://www.omg.org/spec/OCL/2.4}}.
\newblock


\bibitem[Standard(2024)]%
        {alfa}
\bibfield{author}{\bibinfo{person}{OASIS Standard}.}
  \bibinfo{year}{2024}\natexlab{}.
\newblock \bibinfo{title}{{ALFA} - the Abbreviated Language for Authorization}.
\newblock \bibinfo{howpublished}{\url{https://alfa.guide/}}.
\newblock


\bibitem[Tekle and Liu(2010)]%
        {tekle2010precise}
\bibfield{author}{\bibinfo{person}{K~Tuncay Tekle} {and}
  \bibinfo{person}{Yanhong~A Liu}.} \bibinfo{year}{2010}\natexlab{}.
\newblock \showarticletitle{Precise complexity analysis for efficient Datalog
  queries}. In \bibinfo{booktitle}{\emph{Proceedings of the 12th international
  ACM SIGPLAN symposium on Principles and practice of declarative
  programming}}. \bibinfo{pages}{35--44}.
\newblock


\bibitem[Weis(2022)]%
        {policiescode}
\bibfield{author}{\bibinfo{person}{Or Weis}.} \bibinfo{year}{2022}\natexlab{}.
\newblock \bibinfo{title}{What is Policy as Code?}
\newblock
  \bibinfo{howpublished}{\url{https://www.permit.io/blog/what-is-policy-as-code}}.
\newblock


\bibitem[Yang et~al\mbox{.}(2011)]%
        {yang2011finding}
\bibfield{author}{\bibinfo{person}{Xuejun Yang}, \bibinfo{person}{Yang Chen},
  \bibinfo{person}{Eric Eide}, {and} \bibinfo{person}{John Regehr}.}
  \bibinfo{year}{2011}\natexlab{}.
\newblock \showarticletitle{Finding and understanding bugs in C compilers}. In
  \bibinfo{booktitle}{\emph{Proceedings of the 32nd ACM SIGPLAN conference on
  Programming language design and implementation}}. \bibinfo{pages}{283--294}.
\newblock


\end{thebibliography}

\iftoggle{extended}{
\newpage
\appendix
\section{Authorization models for example applications}
\label{sec:code-appendix}

This appendix presents the full authorization models for our \gdrive, \github, and TinyTodo examples, in \cedar, \fga, and \rego.

\subsection{\gdrive}

\subsubsection{\cedar}

\cedar schema for \gdrive:
\begin{schemablock}
entity User in [Group] {
    documentsAndFoldersWithViewAccess: View,
    ownedDocuments: Set<Document>,
    ownedFolders: Set<Folder>,
}
entity Document in [Folder, View] {
    isPublic: Boolean,
};
entity Group;
entity Folder in [Folder, View];
entity View in [View];
action readDocument, writeDocument, shareDocument, changeDocumentOwner
    appliesTo { principal: [User], resource: [Document] };
action createDocumentInFolder
    appliesTo { principal: [User], resource: [Folder] };
\end{schemablock}

\cedar policies for \gdrive:

\noindent
\begin{tabular}{ll}
\begin{minipage}{0.5\textwidth}
\begin{cedarblock}
// If a user is granted view access to a document
// or a parent folder, or the enclosing group is
// granted view access, then they can read the
// document.
permit (
    principal,
    action == Action::"readDocument",
    resource
)
when {
    resource in principal.documentsAndFoldersWithViewAccess
};

// A document's owner (or owners of the parent
// folder) can read, write to, or share the
// document.
permit (
    principal,
    action in [
        Action::"readDocument",
        Action::"writeDocument",
        Action::"shareDocument"
    ],
    resource
)
when
{
    resource in principal.ownedDocuments
    || resource in principal.ownedFolders
};
\end{cedarblock}
\end{minipage} &
\begin{minipage}{0.47\textwidth}
\begin{cedarblock}
// A document's owner can change the owner.
permit (
    principal,
    action == Action::"changeDocumentOwner",
    resource
)
when {
    principal.ownedDocuments.contains(resource)
};

// A folder's owner can create documents.
permit (
    principal,
    action == Action::"createDocumentInFolder",
    resource
)
when {
    principal.ownedFolders.contains(resource)
};

// If a document is public, anyone can read it.
permit (
    principal,
    action == Action::"readDocument",
    resource
)
when { resource.isPublic };
\end{cedarblock}
\end{minipage} \\
\end{tabular}

\subsubsection{\cedar templates encoding}

\cedar schema for \gdrive with templates:
\begin{schemablock}
entity User in [Group] {
  ownedDocuments: Set<Document>,
  ownedFolders: Set<Folder>
};
entity Document in [Folder] { isPublic: Boolean };
entity Group;
entity Folder in [Folder];
action readDocument, writeDocument, shareDocument, changeDocumentOwner
  appliesTo { principal: [User], resource: [Document] };
action createDocumentInFolder
  appliesTo { principal: [User], resource: [Folder] };
\end{schemablock}

\cedar policies for \gdrive with templates: all are unchanged from the policies above, except the first policy becomes the following template:
\begin{cedarblock}
// Template to grant a user or group read access to a document or folder.
permit (
    principal in ?principal,
    action == Action::"readDocument",
    resource in ?resource
);
\end{cedarblock}

In the templates encoding, to grant a user or group read access to a document or folder, instead of adding the document or folder to the appropriate \code{View} parent, we instantiate this template with the appropriate principal and resource.

\subsubsection{\fga}

\fga authorization model for \gdrive:
(Note that this is slightly tweaked from \fga's provided version, to allow write and share permissions for owners of transitive parent Folder(s) and not just the direct parent Folder(s), as noted in the footnote in \Cref{sec:eval:gdrive-app}.)
\begin{fgablock}
model
  schema 1.1
type user
type group
  relations
    define member: [user]
type folder
  relations
    define can_create_file: owner
    define can_write_files: owner or can_write_files from parent
    define can_share_files: owner or can_share_files from parent
    define owner: [user]
    define parent: [folder]
    define viewer: [user,user:*,group#member] or owner or viewer from parent
type doc
  relations
    define owner: [user]
    define parent: [folder]
    define viewer: [user,user:*,group#member]
    define can_change_owner: owner
    define can_read: viewer or owner or viewer from parent
    define can_write: owner or can_write_files from parent
    define can_share: owner or can_share_files from parent
\end{fgablock}

\subsubsection{\rego}

\rego policies for \gdrive:
\begin{regoblock}
package gdrive
import future.keywords.in

################################################################################
#### GDrive Rego Policy
#### input -> dict:
####   principal -> dict:
####   	documentAndFolderWithViewAccess : list of string
####   	ownedFolders : list of string
####   	ownedDocuments : list of string
####   action -> string
####   resource -> dict:
####    uid : string
####	isPublic : bool
####   files : file-graph
################################################################################

transitive_actions := ["Action::\"readDocument\"", "Action::\"writeDocument\"", "Action::\"shareDocument\""]

## Any public resource can be read
allow {
    input.action == "Action::\"readDocument\""
    input.resource.isPublic
}

## Any file reachable from our view-set can be read
allow {
    input.action == "Action::\"readDocument\""
    input.resource.uid in graph.reachable(input.files, input.principal.documentsAndFoldersWithViewAccess)
}

# Any file we own can be read/write/share-d
allow {
    input.action in transitive_actions
    input.resource.uid in input.principal.ownedDocuments
}

# Any file contains in a folder we own can be read/write/share-d
allow {
    input.action in transitive_actions
    input.principal.ownedFolders[_] in graph.reachable(input.files, [input.resource.uid])

}

# Any document we own we can change the owner of
allow {
    input.action == "Action::\"changeDocumentOwner\""
    input.resource.uid in input.principal.ownedDocuments
}

# Any folder we own we can create a document in
allow {
    input.action == "Action::\"createDocumentInFolder\""
    input.resource.uid in input.principal.ownedFolders
}
\end{regoblock}

\subsection{\github}

\subsubsection{\cedar}

\cedar schema for \github:
\begin{schemablock}
entity User in [Organization, Team, OrgPermission, RepoPermission] {}
entity Organization in [Organization, OrgPermission] {
    admins: OrgPermission,
    writers: OrgPermission,
    readers: OrgPermission,
};
entity Team in [Team, RepoPermission];
entity Repository {
    admins: RepoPermission,
    maintainers: RepoPermission,
    writers: RepoPermission,
    triagers: RepoPermission,
    readers: RepoPermission,
}
entity OrgPermission in [OrgPermission]
entity RepoPermission in [RepoPermission]
action readRepository in [triageRepository]
    appliesTo { principal: [User], resource: [Repository] };
action triageRepository in [writeRepository]
    appliesTo { principal: [User], resource: [Repository] };
action writeRepository in [maintainRepository]
    appliesTo { principal: [User], resource: [Repository] };
action maintainRepository in [administrateRepository]
    appliesTo { principal: [User], resource: [Repository] };
action admin
    appliesTo { principal: [User], resource: [Repository] };
\end{schemablock}

In this schema, each action is both a group and an action, which makes it easy to ensure that (e.g.) all writers also have read permissions, using \code{action in} as shown below.

\cedar policies for \github:

\noindent
\begin{tabular}{ll}
\begin{minipage}{0.48\textwidth}
\begin{cedarblock}
// Users can perform an action if they have been
// granted the necessary permission.
permit (
    principal,
    action == Action::"readRepository",
    resource
)
when { principal in resource.readers };

permit (
    principal,
    action in Action::"triageRepository",
    resource
)
when { principal in resource.triagers };

permit (
    principal,
    action in Action::"writeRepository",
    resource
)
when { principal in resource.writers };

permit (
    principal,
    action in Action::"maintainRepository",
    resource
)
when { principal in resource.maintainers };

permit (
    principal,
    action in Action::"administrateRepository",
    resource
)
when { principal in resource.admins };
\end{cedarblock}
\end{minipage} &
\begin{minipage}{0.48\textwidth}
\begin{cedarblock}
// Users also inherit permissions from the
// owner of a repository.
permit (
    principal,
    action == Action::"readRepository",
    resource
)
when { principal in resource.owner.readers };

permit (
    principal,
    action in Action::"writeRepository",
    resource
)
when { principal in resource.owner.writers };

permit (
    principal,
    action in Action::"administrateRepository",
    resource
)
when { principal in resource.owner.admins };
\end{cedarblock}
\end{minipage} \\
\end{tabular}

\subsubsection{\cedar templates encoding}

\cedar schema for \github with templates:
\begin{schemablock}
entity User in [Organization, Team, OrgPermission];
entity Organization in [OrgPermission] {
  admins: OrgPermission,
  readers: OrgPermission,
  writers: OrgPermission
};
entity Team in [Team];
entity Repository { owner: Organization };
entity OrgPermission;
action readRepository in [triageRepository]
    appliesTo { principal: [User], resource: [Repository] };
action triageRepository in [writeRepository]
    appliesTo { principal: [User], resource: [Repository] };
action writeRepository in [maintainRepository]
    appliesTo { principal: [User], resource: [Repository] };
action maintainRepository in [administrateRepository]
    appliesTo { principal: [User], resource: [Repository] };
action admin
    appliesTo { principal: [User], resource: [Repository] };
\end{schemablock}

\cedar policies for \github with templates: (the policies on the right are unchanged from above)

\noindent
\begin{tabular}{ll}
\begin{minipage}{0.48\textwidth}
\begin{cedarblock}
// Grant read access to a user, team, or organization.
@id("readTemplate")
permit (
    principal in ?principal,
    action == Action::"readRepository",
    resource == ?resource
);

// Grant triage access to a user, team, or organization.
@id("triageTemplate")
permit (
    principal in ?principal,
    action in Action::"triageRepository",
    resource == ?resource
);

// Grant write access to a user, team, or organization.
@id("writeTemplate")
permit (
    principal in ?principal,
    action in Action::"writeRepository",
    resource == ?resource
);

// Grant maintainer access to a user, team, or organization.
@id("maintainTemplate")
permit (
    principal in ?principal,
    action in Action::"maintainRepository",
    resource == ?resource
);

// Grant admin access to a user, team, or organization.
@id("adminTemplate")
permit (
    principal in ?principal,
    action in Action::"administrateRepository",
    resource == ?resource
);
\end{cedarblock}
\end{minipage} &
\begin{minipage}{0.48\textwidth}
\begin{cedarblock}
// Users also inherit permissions from the
// owner of a repository.
permit (
    principal,
    action == Action::"readRepository",
    resource
)
when { principal in resource.owner.readers };

permit (
    principal,
    action in Action::"writeRepository",
    resource
)
when { principal in resource.owner.writers };

permit (
    principal,
    action in Action::"administrateRepository",
    resource
)
when { principal in resource.owner.admins };
\end{cedarblock}
\end{minipage} \\
\end{tabular}

\subsubsection{\fga}

\fga authorization model for \github:
\begin{fgablock}
model
  schema 1.1
type user
type team
  relations
    define member: [user,team#member]
type repo
  relations
    define admin: [user,team#member] or repo_admin from owner
    define maintainer: [user,team#member] or admin
    define owner: [organization]
    define reader: [user,team#member] or triager or repo_reader from owner
    define triager: [user,team#member] or writer
    define writer: [user,team#member] or maintainer or repo_writer from owner
type organization
  relations
    define member: [user] or owner
    define owner: [user]
    define repo_admin: [user,organization#member]
    define repo_reader: [user,organization#member]
    define repo_writer: [user,organization#member]
\end{fgablock}

\subsubsection{\rego}

\rego policies for \github:
\begin{regoblock}
package github
import future.keywords.in


################################################################################
#### Github Rego Policy
#### input -> dict:
####   principal -> string
####   action -> string
####   resource -> dict:
####     readers -> string
####     writers -> string
####     admins -> string
####     triagers -> string
####     maintainers -> string
####     owner -> dict:
####     	readers -> string
####     	writers -> string
####     	admins -> string
####   orgs -> dict (Org-Graph)
################################################################################


actions = {
    "Action::\"readRepository\"":
        ["Action::\"triageRepository\"", "Action::\"writeRepository\"", "Action::\"maintainRepository\"", "Action::\"administrateRepository\""],
    "Action::\"triageRepository\"":
        ["Action::\"writeRepository\"", "Action::\"maintainRepository\"", "Action::\"administrateRepository\""],
    "Action::\"writeRepository\"":
        ["Action::\"maintainRepository\"", "Action::\"administrateRepository\""],
    "Action::\"maintainRepository\"":
        ["Action::\"administrateRepository\""],
    "Action::\"administrateRepository\"": [],
}

## Look up applicable actions (ex: 'write' -> 'read', 'triage', 'write')
applicable = graph.reachable(actions, [input.action])

## Can a specific action be performed
can_perform[action] = true {
    ## Look up the repo's object capability for this action
    cap := input.resource[action]

    ## Can we reach that capability in the org chart?
    cap in graph.reachable(input.orgs, [input.principal])
}

can_perform[action] = true  {
    ## Do the same thing, but user the owner's object capability
    cap := input.resource.owner[action]

    ## Can we reach that capability in the org chart?
    cap in graph.reachable(input.orgs, [input.principal])
}


## We are authorized if any of the applicable actions are allowed
allow = true {
    some action
    action in applicable
    can_perform[action]
}
\end{regoblock}

\subsection{TinyTodo}

\subsubsection{\cedar}

\cedar schema for TinyTodo:
\begin{schemablock}
entity Application;
entity User in [Team, Application] { name: String };
entity List in [Application] {
  editors: Team,
  name: String,
  owner: User,
  readers: Team,
  tasks: Set<{ id: Long, name: String, state: String }>,
};
entity Team in [Team, Application];
action CreateList, GetLists
  appliesTo { principal: [User], resource: [Application] };
action EditShares, UpdateTask, CreateTask, GetList, UpdateList, DeleteTask, DeleteList
  appliesTo { principal: [User], resource: [List] };
\end{schemablock}

\cedar policies for TinyTodo: (somewhat different from those shown in \Cref{fig:tinytodo-policies}, which were modified for exposition)

\noindent
\begin{tabular}{ll}
\begin{minipage}{.5\textwidth}
\begin{cedarblock}
// Policy 0: Any User can create a list
// and see what lists they own.
permit(
    principal,
    action in [Action::"CreateList",
               Action::"GetOwnedLists"],
    resource == Application::"TinyTodo");

// Policy 2: A User can see a List if they are
// either a reader or editor.
permit (
    principal,
    action == Action::"GetList",
    resource
)
when {
    principal in resource.readers ||
    principal in resource.editors
};
\end{cedarblock}
\end{minipage} &
\begin{minipage}{.47\textwidth}
\begin{cedarblock}
// Policy 1: Any User can perform any action
// on a List they own.
permit(principal, action, resource)
when {
    resource is List &&
    resource.owner == principal
};

// Policy 3: A User can update a List and its
// tasks if they are an editor
permit (
    principal,
    action in
        [Action::"UpdateList",
         Action::"CreateTask",
         Action::"UpdateTask",
         Action::"DeleteTask"],
    resource
)
when {
    principal in resource.editors
};
\end{cedarblock}
\end{minipage} \\
\end{tabular}

\subsubsection{\fga}

\fga authorization model for TinyTodo:
\begin{fgablock}
model
  schema 1.1
type application
  relations
    define member: [user,team#member,user:*]
    define create_list: member
    define get_lists: member
type user
type team
  relations
    define member: [user,team#member]
type list
  relations
    define owner: [user]
    define editor: [user,team#member] or owner
    define reader: [user,team#member,user:*] or owner or editor
    define get_list: reader
    define update_list: editor
    define create_task: editor
    define update_task: editor
    define delete_task: editor
    define edit_shares: owner
    define delete_list: owner
\end{fgablock}

\subsubsection{\rego}

\rego policies for TinyTodo:
\begin{regoblock}
package tinytodo
import future.keywords.in

################################################################################
###### Static Action Data
################################################################################

public_actions := ["Action::\"CreateList\"", "Action::\"GetLists\""]
read_actions := ["Action::\"GetList\""]
write_actions := ["Action::\"UpdateList\"", "Action::\"CreateTask\"", "Action::\"UpdateTask\"", "Action::\"DeleteTask\""]


################################################################################
###### Read/Write rules
################################################################################


# Check if we're allowed to write
allow_write = true
# We can write if we are directly in the writers list
{
    some user in input.Request.Resource.Writers
    input.Request.Principal == user
}
# Or if we are transitively in a group in the writers list
{
    some group in input.Request.Resource.Writers
    group in graph.reachable(input.Data.Groups, [input.Request.Principal])
}

# Check if we're allowed to read
allow_read = true
# We can read if we are directly in the readers list
{
    some user in input.Request.Resource.Readers
    input.Request.Principal == user
}
# Or if we are transitively in a group in the readers list
{
    some group in input.Request.Resource.Readers
    group in graph.reachable(input.Data.Groups, [input.Request.Principal])
}
# Or if we are allowed to write
{
    allow_write
}


################################################################################
###### Toplevel rules
################################################################################


# This tracks policy violations
default violations = set()

# Any user can take public actions
allow = true {
    input.Request.Action in public_actions
}

# Any user can take actions on resources they own
allow = true {
    input.Request.Resource.Owner == input.Request.Principal
}

# Read actions are allowed if the user can read
allow = true {
    read_actions[_] = input.Request.Action
    allow_read
}

# Write actions are allowed if the user can write
allow = true {
    write_actions[_] = input.Request.Action
    allow_write
}
\end{regoblock}

}{}

\end{document}